\documentclass{article}
\usepackage[style=apa,backend=biber, uniquename=false]{biblatex}
\usepackage{amsmath}
\usepackage{amssymb}
\usepackage{setspace}
\usepackage[affil-it]{authblk}
\usepackage{subcaption}
\usepackage{xcolor}
\usepackage{indentfirst}
\usepackage[a4paper, portrait, margin=1.25in]{geometry}
\usepackage{float}
\usepackage{graphicx} 
\graphicspath{ {./images/} }

\title{In the Shadow of Silence: Modelling Missing Data in the Dark Networks of Crime and Terrorists}
\author[1]{Jonathan Januar \footnote{ Corresponding author \\   \textit{E-mail address}: jonathan.januar@unimelb.edu.au (J. Januar).}}
\author[2]{H Colin Gallagher}
\author[1]{Johan Koskinen}
\affil[1]{Melbourne School of Psychological Sciences, University of Melbourne, Australia}
\affil[2]{Melbourne School of Population and Global Health, University of Melbourne, Australia}

\date{}

\begin{document}

\maketitle

\begin{abstract} 	  
The clandestine nature of covert networks makes reliable data difficult to obtain and leads to concerns with missing data. We explore the use of network models to represent missingness mechanisms. Exponential random graph models provide a flexible way of parameterising departures from conventional missingness assumptions and data management practices. We demonstrate the effects of model specification, true network structure, and different not-at-random missingness mechanisms across six empirical covert networks. Our framework for modelling realistic missingness mechanisms investigates potential inferential pitfalls, evaluates decisions in collecting data, and offers the opportunity to incorporate non-random missingness into the estimation of network generating mechanisms.

\end{abstract}

\break

\section{Introduction}

When encountering missing data, network analysts are often uncertain of methods to understand and deal with the missingness. While missing data is widely known to be a complicating issue across a variety of network methods, the consequences of missing data are as complicated, disparate, and context-dependent as the processes that generate such missingness. Biases, statistical or otherwise, incurred in estimation procedures for network models may be exacerbated when measurement of the network is itself a difficult and potentially biased procedure. Such data collection environments are commonplace for covert networks as direct observation of the networks' actors and tie variables are difficult. It can be especially difficult to understand whether covert networks have certain structures due to underlying social processes, or as an artefact of the method (or methods) by which the network was observed and interpreted. 

We propose the use of an explicit statistical model for missingness to describe various empirical biases encountered when constructing covert networks. We do so by statistically formulating assumptions of how tie variables are systematically missing. Since missingness indicators are structured in a way that is mathematically identical to adjacency matrices, missing tie-variables can be modelled using statistical network models. In particular, we first posit that (i) missing data are not evenly distributed in the network; (ii) the probability of dyads being missing are dependent through the nodes; (iii) the likelihood of a tie-variable being missing cannot reasonably be assumed to be independent of the network structure. Secondly, we propose that the exponential random graph model (ERGMs; Frank \& Strauss, 1986; Snijders et al., 2006), normally used to model tie-variables, is able to address (i) through (iii) when used as a model for the missingness indicators.

While the method is applicable to a variety of network settings, we aim to specifically address systematic patterns in the sampling and missingness of covert network data. We consider the network analyst point of view - be they a researcher or a practitioner - and demonstrate the consequences for inference of common decisions made by a network analyst when they handle missing covert network data.

\section{Covert networks}

Covert (or 'dark' or 'illicit') networks are social networks of individuals that operate in a concealed context. While covert networks do not necessarily refer to networks that maintain illicit or illegal activities (e.g., men who have sex with men, clubbers, persecuted minorities) the term is often used to refer to networks involved in criminal or terrorist activity \parencite{diviak_criminal_2020}.  Whether `covert' refers to the actors (the nodes) or their relational activities (the ties) is not made clear in the literature, largely due to the use of networks as an abstract concept and not as a framework for theory and methods \parencite{robins_multilevel_2023, stys_brokering_2020}. Nonetheless, a vital aspect in the definition of how a covert network is defined is that they are always involved in secrecy. 

The empirical observation of these networks is therefore fundamentally at odds with their operation; secrecy can relate to either membership or affiliation or activity. However, regardless of the exact nature of what is kept secret, it is evident that trustworthy covert network data are especially difficult to collect. Covert network data are obviously not exclusively the domain of academics. Network approaches are becoming increasingly important in criminal and military intelligence work as well as in applied policing and law enforcement. Arguably, the stakes are even higher when acting on faulty and flawed network data when enforcing the law than when explaining criminal networks.

Traditional data collection methods such as the network survey (name generators, rosters, etc) often cannot be administered in covert settings with the expectation of genuine responses. Instead, data collection may rely on multiple overlapping regimes of active surveillance and passive observation, with the observed network also subject to multilayered censoring as a result of policing decisions, legal judgements, journalistic practices, and so forth \parencite{bright_using_2021, berlusconi_all_2013, campana_listening_2012}. Therefore, covert networks may be networks with potentially interesting properties as a result of their context; however, this context includes not only covert social activity itself, but also the methods by which affiliations were observed, censored, and interpreted by multiple observers. The construction and analysis of covert networks therefore carries additional uncertainty when conventional SNA concepts are applied to them.

Beginning with the \textcite{erickson_secret_1981} paper that constructed secret society networks, research involving covert networks have constructed networks out of public records \parencite{krebs_mapping_2002}, wiretaps \parencite{baker_social_1993}, associative ties from surveillance \parencite{coutinho_multilevel_2020}, or constructed with help of informants \parencite{morselli_inside_2009}. However, methodological concerns have repeatedly been brought up regarding the aforementioned secondhand data sources used to construct covert networks \parencite{faust_social_2019, campana_studying_2022}. Among these concerns, the most pertinent issues relate to data validity and missing data \parencite{bright_using_2021}. Unlike network studies that survey a population, covert network data is rarely directly measured. The consequence of all these issues result in improperly defined nodes, ties, and network boundary which are greatly detrimental to the capability of SNA to make inferences regarding structural tendencies of the network.

Despite these concerns, the biases in constructing covert networks have not been well researched. Neither is the effect of compounding these biases in multiple observation processes that then result in a constructed network. There are biases in how easily observable certain members of the covert network are which can introduce biases if the constructed network only contained observed (e.g., arrested) individuals. To name a few, there may be biases in the observed data if the same target network had multiple organisations surveilling them \parencite{berlusconi_come_2022}. There may be biases in clarifying membership to the network as illicit actors in a covert network can have legitimate ties to non-illicit actors outside of the covert network \parencite{bouchard_collaboration_2020, stys_brokering_2020}.  Covert network researchers may also compile relational information from multiple sources which invite the potential of composite tie definitions and the same actor being recorded multiple times \parencite{bright_using_2021, carrington_co-offending_2009}. The use of an ill-defined co-associative tie definition generally invites misinterpretation from anyone involved in the observation process \parencite{faust_social_2019}. Ultimately, the data used to construct covert networks can be the result of a convoluted observation process where biases and uncertainties in the data can be compounded at every step to obfuscate the constructed network the researcher analyses. A key factor in the occurrence of these issues is the process by which the data are observed or conversely, the process by which the data are missing.

We constrict the definition of a covert network to two empirical guided elements. Firstly, covert networks are networks that cannot be directly surveyed in a traditional SNA sense due to their secretive nature. By relying on ambiguous data sources for generic co-conspiracy tie definitions, we assume that undirected edges are more [common/salient] choice for representing covert ties than directed arcs \parencite{bright_using_2021}. While directed covert networks can be constructed, directed ties imply some form of agency on the nodes. Implying agency would certainly be theoretically valuable for covert settings. However, observing the intentions of individuals in covert settings is practically very difficult as the actors' intentionality would need to be inferred. As such, we are left with the observation of a covert activity (i.e., a crime has occurred) and the joint production thereof (the people who are involved). Hence, favouring undirected ties is a practical matter. 

Secondly, the secretive context of a covert network is very important in explaining the actions of the actors but does not translate to expected network structures. For example, while often said to be comparatively sparser than non-covert networks \parencite{krebs_mapping_2002}, covert networks in general are not distinctively sparser or denser than non-covert social networks \parencite{oliver_covert_2014}.

\subsection{Missingness and data gathering in covert networks}

Missing data in networks involve unique difficulties when compared to missing data in non-network (independent) data, but the nature of covert networks introduces even more complications. Missingness encountered in network analysis is often categorised into missing nodes and/or missing ties \parencite{kossinets_effects_2006, huisman_imputation_2009, koskinen_bayesian_2013}. Missing nodes refer to completely missing individuals in the network, while missing ties refer to an identified pair of individuals with a missing tie-indicator.  A common source of missingness for non-covert social network studies are non-respondents, which refer to individuals who do not respond to surveys and consequently have completely missing outgoing ties. 

However, missingness in covert networks is different when compared to missingness in more typical network studies where conventional data collection methods can be deployed. For example, non-respondents as defined above do not truly exist for covert networks, as a survey is typically not used and there are consequently no explicit respondents. We may instead view the network's boundary specification as an activity- or event-based network (e.g., all individuals who participate in an activity; \cite{laumann_boundary_1983}) rather than an attribute-based network. Therefore, an equivalent to the non-respondent could be the result of insufficient observation of the activity that defines the network in the first place.

Covert networks present especially difficult problems when it comes to the specification of a network boundary. A covert actor will likely have non-covert social relationships outside of their covert network. Additionally, overlapping networks where the same individuals are in the covert and some non-covert network also presents a challenge in identifying the members of the covert network. Generally, these 'fuzzy' boundaries occur when there are ambiguities in the definition of a covert network boundary \parencite{stys_brokering_2020, bouchard_collaboration_2020, burcher_social_2015}. This results in challenges of determining which actors are a part of the covert network. In general, when a researcher or analyst has to construct a network, missingness can manifest in the many different steps that lead to the production of the information used by the researcher or analyst. Broadly speaking, it depends on the type of information the researcher has access to (e.g., public records, wiretap transcripts, auditory recordings, consulting informants, etc.). 

A case in point for a layered network construction process is in terms of the standard taxonomy of intelligence gathering and how this is used to construct covert networks. The typical taxonomy is referred to as different \emph{intelligence collection disciplines}, comprising open source intelligence (OSINT),
human intelligence (HUMINT),
signals intelligence (SIGINT),
geospatial intelligence (GEOINT),
and measurement and signature intelligence (MASINT)
\parencite[see, e.g.][]{henrico_intelligence_2024}.  HUMINT describes intelligence derived from information collected and provided by human sources \parencite{pigeon_humint_2002}. SIGINT refers to intelligence gathered from communication or electronic devices. Lastly, OSINT refers to information available from open sources.

Academic researchers often draw on intelligence data provided by intelligence gathering agencies, sometimes of a specific collection discipline, such as SIGINT \parencite{henrico_intelligence_2024}, at other times there data are derived from a combination of disciplines, such as \textcite{coutinho_multilevel_2020}, whose data stem from a combination of HUMINT, SIGINT, and OSINT, which were triangulated to select specific subsets of co-operation between criminals in the process of operating their criminal enterprise. 

With the ethical and resource constraint that researchers face, if they are not provided data from government agencies, they resort to a form of OSINT. For example, 
\textcite{rhodes_use_2011} could be said to draw on OSINT when they construct a network of a Greek terrorist organisation following public trials which released data about their personal attributes, their activities in relation to the operation of the terrorist group, their leadership roles within the organisation, and membership of three different factions within the group. The open source information may itself have been based on extensive investigations using SIGINT, HUMINT, etc. When intelligence reports are gathered by the police, criminal, or military intelligence, selecting ties through the subsets available in the intelligence reports make it less likely to include less specific tie types (e.g., romantic, familial, etc.). 

Different categories of intelligence may be susceptible to specific biases. OSINT may not reflect more secretive aspects of the covert network as the information is dependent on which aspects of the network were observable. HUMINT and SIGINT may also be biased from the process by which they were obtained. For example, depending on the stage of their investigation, police investigations and reports may have their own biases for the potential individuals relevant to the network. 

The investigators, or broadly speaking any other individual involved in the process of constructing a network, may also be susceptible to cognitive biases, including various confirmation biases. An example of this bias can be seen as a 'spotlight' effect where some members of the covert network are featured more prominently than others, resulting in a non-random distribution of missing ties \parencite{smith_trust_2016}. More broadly, a recent review of cognitive biases in criminal case evaluation \parencite{meterko_cognitive_2022} detail multiple studies evaluating a variety of cognitive biases identified by researchers and law enforcement personnel.

Among the identified cognitive biases described in \textcite{meterko_cognitive_2022} are those that directly relate to personal evaluations of information. These include a tendency to overestimate the validity of partial information \parencite{ditrich_cognitive_2015}, where police officers' initial belief about the innocence or guilt about a suspect in a fictional case predicts their subsequent evaluation of ambiguous evidence, which subsequently predicted their final belief about the suspect's innocence or guilt \parencite{charman_cognitive_2017, ask_elasticity_2008}. 

Another tendency was the detection of the presence of an attribute more than its absence (i.e., a feature positivity effect). In an experiment, \textcite{eerland_out_2012} found that the presence of fingerprints was more easily remembered and used to make decisions about a criminal case than an absence. In summary, data gathering and missingness is a highly layered and convoluted process in covert networks and there is a lack of understanding of how statistical network models perform under these dire data collection environments.

An avenue to introduce more complexity in the generation of missingness mechanisms is to explicitly specify a model for the missingness. In this situation, the 'model' would be a statistical model that is assumed to capture some of the systematic missingness induced by the convoluted observation process used to construct a network. This model would capture biases in observations and would weigh the probability of missing tie variables depending on how the model is specified. Previous studies that have researched the impact of missing data on network analysis often rely on simpler missingness assumptions \parencite{krause_missing_2018, huisman_imputation_2009, smith_structural_2013}, and while some studies have simulated missingness mechanisms that may depend on the observed data \parencite{smith_network_2017, smith_network_2022}, explicitly describing a model for the missingness mechanism has not been attempted. There is a notable exception in \textcite{koskinen_bayesian_2019}, where an explicit sampling mechanism is specified in the model, however it is different as the observation process we seek to define is more elaborate. We define the missingness model as a statistical model to describe the assumptions about the process by which missing data is encountered.

The current study has two aims. The first aim is to propose the use of statistical models for missingness as a framework to represent the processes that generate missing network data. The second aim is to review some models for missingness,  examine the assumptions that these models make for the missingness that occurs, and investigating systematic effects on conclusions drawn on incomplete network data. These studies will inform us not only about what potential errors we run the risk of making under plausible missingness scenarios but also pave the way for incorporating and accounting for these mechanisms in model estimation in the future.

We will find, for example, that missingness generated as a function of different levels of conditioning on the network can have drastically different observed network statistics. We also find the structures of the true network to affect both the estimation process and the specific biases incurred from multiple missingness mechanisms. Lastly, we find the specification of the estimation model to greatly alleviate model convergence difficulties, but does not noticeably mitigate the biased estimated parameters.

Our aim is not to identify which missing data mechanisms operate in covert settings, which would require a more ethnographic approach. Nor is our aim to give a definitive answer on how much missingness may be tolerated in our statistical analyses. Instead, our aim is to provide a formalised mathematical and statistical approach to modeling the impact of missing data, providing the analytical foundations to operationalise more realistic missingness assumptions, and test out the impacts of these assumptions on common empirical covert network research.

\section{Notation and definitions}

Before describing further details of what an explicit model for the missingness is, we will first define how network data are represented in mathematical objects and how missingness can be represented. If we first start from the construction of a network, we can collect all the identified actors in a list. From the identified actors, we can then start mapping relationships (e.g., when certain actors were arrested together or any other possible tie definition) and storing them in a list. The list of actors can be represented by a set of vertices $V$ and the relationships in the edge list can be represented by a set of edges $E$. When combining the vertices and edges, we have a graph $G(V,E)$. However, representing the constructed network as a graph implicitly assumes that there are no missing actors. The vertex set $V$ is assumed to be fixed for the graph $G(V,E)$ and there cannot be any missing actors in $V$ which are strong assumptions to make of a constructed network as it assumes the researcher has confidently identified all the relevant actors of the covert network. We will not be exploring missing actors\footnote{This is obviously both a crucial issue, not only in covert networks but in all manner of network research, as well as an intractable problem. See \textcite{koskinen_bayesian_2013} for an extensive discussion of the boundary specification issue \parencite{laumann_boundary_1983} in the context of `missing actors'}, but the vertex set $V$ being fixed is an assumption that is important to emphasise as it describes a situation where missingness can have large impact on the constructed network.

Once we have assumed that vertex set $V$ contained all the relevant actors, we can represent the graph $G(V,E)$ as an adjacency matrix $\mathbf{X}$. If vertex set $V$ had $n$ elements, $\mathbf{X}$ is an $n \times n$ binary matrix representing the edges in set $E$. The adjacency matrix $\mathbf{X}$ contains $n(n-1)$ (random) tie variables $x_{ij}$ indexed by $\{i,j\} \in \{1, 2, ..., n\}$. Index $i$ represents the row index and $j$ represents the column index. For an undirected graph, tie variables $x_{ij}$ are defined as

$$ x_{ij} = 
\begin{cases} \ 0 & \text{if edge } \{i,j\} \notin E \ \text{or } i = j \\ 
\ 1 \ & \text{if edge } \{i,j\} \in E \end{cases}. $$

When missingness of tie variables is involved, the true value of the tie variable is not observed in the collected data. This implies that the researcher cannot assume the missing tie variables to either be 0 or 1 because the true value of the variable is unknown.

\noindent In subsequent descriptions of the observed adjacency matrix in this manuscript, we take this distinction to be apparent from the context. One way to represent missing network data is with a missing data indicator.
 This missing data indicator indicates whether the possible data points in the data are missing or not. A single indicator can be defined as 

$$d = 
\begin{cases} \ 0 & \text{if variable } x \ \text{has been observed} \\
\ 1 \ & \text{if variable } x \ \text{has not been observed}
\end{cases}.$$

We note here that the missingness indicator can easily be re-binarised to represent the sampling of data instead. Since the data are an adjacency matrix $\mathbf{X}$, with some missingness, the indicators can be collected into a matrix $\mathbf{D}$. Therefore, the missing data indicator matrix $\mathbf{D}$ would also be an $n \times n$ binary matrix representing missingness for each corresponding tie variable. Similar to $\mathbf{X}$, $\mathbf{D}$ also contains $n(n-1)$ missing tie variable indicators $d_{ij}$ indexed with indices $\{i, j \}, \in \{1, 2, ..., n\}$ equivalent to the tie variables $x_{ij}$ in the observed adjacency matrix $\mathbf{X}^*$. For an undirected graph, missing indicators $d_{ij}$ are defined as

$$d_{ij} = 
\begin{cases} \ 0 & \text{if tie variable } x_{ij} \ \text{has been observed or } i = j \\
\ 1 \ & \text{if tie variable } x_{ij} \ \text{has not been observed}
\end{cases}.$$

With $\mathbf{D}$ defined, we make the distinction here between the adjacency matrix $\mathbf{X}$ and the observed adjacency matrix $\mathbf{X}^*$. $\mathbf{X}^*$ is defined as $\mathbf{X}$ with the missing elements replaced by $NA$. the observed tie variables $x^{\ast}_{ij} \in \mathbf{X}$ can alternatively be defined as

\begin{equation}\label{def:xobsmiss}
x^{\ast}_{ij} = 
\begin{cases} 
\ 0 & \text{if } x_{ij} = 0 \text{ and } \ d_{ij} = 0 \\ 
\ 1 & \text{if } x_{ij} = 1 \ \text{ and }  \ d_{ij} = 0 \\
\ NA & \text{if } d_{ij} = 1
\end{cases}.
\end{equation}

The missing data indicators $\mathbf{D}$ describe which portions of the data have and have not been observed. As seen above, we represent missing values as $NA$, which is common representation in statistical programming languages (e.g., R, Python). A point to note here is that the representation of missing values can be affected by some implicit assumptions. These can be implicit algorithmic assumptions, where missing values may be represented or used in very specific ways depending on the software, or be susceptible to seemingly harmless researcher assumptions. A cautionary example of a seemingly harmless decision would be to `ignore' missing tie-variables and set them all to 0s. We will refer to this a \emph{zero imputation}. This implies the assumption that all missing tie variables are 0, which may be plausible for data where 0 is not a meaningful interpretable value. This does not apply for the adjacency matrix and network data as a whole because 0 represents the absence of a tie (or a 'null tie'). Assuming that all the missing data all represent absent ties is a very strong assumption and is effectively a form of imputation (i.e., null tie imputation as in \cite{huisman_imputation_2009}) rather than a sensible way to represent missing data.

A few notable implicit algorithmic assumptions regarding missingness in popular software can be found in the RSiena, sna, ergm R packages and PNet. In Rsiena, all missing tie variables in the first wave of data is set to 0 as described in the RSiena manual \parencite{snijders_manual_2024}. Some computations in the sna package also tend to ignore missing tie variables and dyads adjacent to any missing tie variables and instead compute the statistics omitting those missing tie variables (e.g., triad census, clustering coefficient). Naturally, graph plotting routines typically set 'NA' to 0. As described in PNet manual \parencite{wang_pnet_2009}, analyses with missing data relies on \textcite{koskinen_bayesian_2013} which assumes missingness is missing at random. The most relevant implicit algorithmic assumption for this paper is the routine for handling missing tie variables in the ergm package, which we elaborate below.

\section{Missingness models}

We begin our model-based representation of missingness by clarifying the similarities of $\mathbf{X}$ and $\mathbf{D}$. Both are square, symmetric, binary matrices with the same range space of $\{0, 1 \}^{V \choose 2}$, so we may be able to apply modelling approaches for adjacency matrix $\mathbf{X}$ to missingness indicator $\mathbf{D}$. If we let $\mathcal{N}={V \choose 2}$, we denote by $\mathcal{D}=\{0, 1 \}^\mathcal{N}$, the range space for the missing tie indicators, and $\mathcal{X}=\{0, 1 \}^\mathcal{N}$, the space of undirected graphs. Since the missing data indicator $\mathbf{D}$ is defined to indicate whether the variables in the collected data are missing or not, the dimensions of $\mathbf{D}$ will always be equal to the mathematical object that represents tie variables. The variables in $\mathbf{D}$ can only ever be binary to note whether a particular tie variable is observed or not and when we speak of networks or simple graphs, the observed tie variables in adjacency matrix $\mathbf{X}$ can also only be binary. Therefore, any model that can describe the tie variables in $\mathbf{X}$ can be used to describe the missing tie variables in $\mathbf{D}$. 

Using a statistical model for $\mathbf{D}$ as a generative model offers flexibility to clarify assumptions about the missingness mechanism. The construction process of a covert network can reflect many different decisions from multiple stakeholders. Biases in the construction process may stem from any stage, from the covert network itself to police investigators observing the network, to researchers accessing police data. There are myriad of potential compounding and interacting biases, meaning that it might be challenging or impossible to reverse-engineer the construction process from the final outcome. In the face of the unknown and layered biases, the use of a statistical model for $\mathbf{D}$ will lend analytical clarity to the systematic biases that affect the probability of a tie variable being missing. The benefit of having a model-based representation is the ability of clarifying specific biases in the construction process through the model specification of the missingness model. Consequently when we specify empirical biases in the model, the missingness model can simulate the outcome of an empirical covert network construction process. We will explore some candidate models to use as missingness models with varying assumptions.

\subsection{Bernoulli random graph models}

The simplest network model we may apply to model $\mathbf{D}$ would be a homogeneous Bernoulli random graph model (hBRGM). Here, the hBRGM would imply that we assume that all indicators $d_{ij}$, are  independent across all dyads and of each tie variable. The probability of a missing tie variable can thus be expressed as, independently and identically for all $\{i,j\}\in \mathcal{N}$

\begin{equation}\label{eqn:hBRGMprob}
\Pr(d_{ij} = 1) = p,
\end{equation}

\noindent where $p$ is the probability of a tie variable being missing that thus applies to all tie variables $d_{ij}$. This model assumes that the probability of any tie variable being missing is completely independent of any other tie variable being missing. This is a very strong assumption indeed, as is assumes that data are missing completely by chance, independent of the covert actors, and not dependent on any property of the data that have been observed or unobserved. While this assumption may be suitable for non-network data observed from experimental observations, it is difficult to justify complete independence of the properties of the data especially with a constructed network, and especially for covert networks.

The probability parameter $p$ does not necessarily need to be completely homogeneous for all the tie variables. Instead of assuming a single parameter $p$ for all the tie variables, we can relax the assumption such that the tie variables are indexed with their own parameter. To be specific, instead of the model (Equation \ref{eqn:hBRGMprob}), for all  $\{i,j\}\in \mathcal{N}$,
we can allow each tie variable $\{i, j \}$ to have its own missing tie variable probability $p_{ij}$, according to the in-homogeneous Bernoulli random graph model (BRGM) 

$$\Pr(d_{ij} = 1) = p_{ij}, \quad \{i,j\}\in \mathcal{N}.$$

\noindent While there may be situations that may suggest homogeneous missing tie probabilities, assuming that missing tie variable probabilities are heterogeneous is a much more flexible assumption. 

There are multiple approaches that can model heterogeneity in missing tie variable probabilities. If either dyad or actor covariates are available, BRGMs can be used to model how the probability of a missing tie variable can depend on covariates. To illustrate an example with actor covariates for an undirected network, let $C$ be an $n \times 1$ vector containing a covariate $c$ for $n$ actors, 

$${\mathrm {logit}}(p_{ij}) = \theta_0 + \theta_1(c_i + c_j),$$

\noindent where $\theta_0$ is an intercept value describing some baseline probability of a missing tie variable, $\theta_1$ is the change in missing tie variable probability dependent on the values of the covariates $c_i$ and $c_j$ for the two nodes $i$ and $j$ in the tie variable $d_{ij}$. An example could be assuming the ages of the $n$ nodes explains the probability of a tie variable being missing. If we assumed that a tie variable involving older individuals are more likely to be missing, the heterogeneous Bernoulli graph would be able to express the heterogeneity of the missing tie variable probabilities as a function of the heterogeneity of node ages.

Heterogeneous Bernoulli graphs as a model for missing tie variable probabilities may be appropriate when the researcher is confident the identified attributes in explaining why certain tie variables are missing. However as the model completely relies on attributes for either the actors or dyads, any missing or irrelevant attributes would limit the usage of this model. Realistically speaking, the heterogeneous Bernoulli model would only be appropriate when the researcher has access to the attribute(s) that sufficiently explain the missing tie variables. This ultimately narrows down the situations where heterogeneous Bernoulli graph are the most preferable models to use to represent missingness.

With undirected edges, we can introduce heterogeneity in a different way using the $\beta$ model \parencite{chatterjee_random_2011} can be used to model the degree distribution. The $\beta$-model is a statistical model for the degree distribution of undirected random graphs which only requires the degree of the nodes to model tie probabilities. For an undirected network with $n$ nodes, the $\beta$-model assumes that an edge between nodes $i$ and $j$ have a probability of

$$\Pr(x_{ij}=1) = \frac{\exp(\beta_i + \beta_j)}{ 1 + \exp(\beta_i + \beta_j)}, 1 \leq i \neq j \leq n,$$

\noindent where $\beta_i$ is an influence parameter for node $i$ to quantify the propensity of node $i$ to have ties. In total, there would be $n$ $\beta$ parameters, one for each node. When $\beta_i$ is large and positive, the degree of node $i$ is expected to be large and when $\beta_i$ is large and negative, the degree of node $i$ is expected to be small. This model assumes that the probability of each tie is dyadically independent conditional on each $\beta$ value.

When applied to missing tie variables, the $\beta$-model offers a model for the counts of missing tie variables for each node in the missingness indicator $\mathbf{D}$. The resulting expression for the $\beta$-model for $\mathbf{D}$ is then

$$p_{ij} =\frac{\exp(\beta_i + \beta_j)}{1 + \exp( \beta_i + \beta_j)}, 1 \leq i \neq j \leq n.$$

\noindent In this case, a large $\beta_i$ would imply that the tie variables $x_{ij}$ of node $i$ are more likely to be missing than for a node $h$ with a smaller $\beta_h$. If $\beta_h$ is very small, then it is more likely that all of tie tie variables involving node $h$ will be observed. The $\beta$ values can thus represent some activities that make some individuals harder to detect than others. The $\beta$-model is ultimately a useful way to model missing tie variables independently conditional on the $\beta$ values for each node. However, at this point the models for $\mathbf{D}$ do not account for any information in the network $\mathbf{X}$.

Another way to incorporate heterogeneity is to identify roles or positions of certain actors within the network and assume that different roles can have different propensities for missing tie variables. Following this line of thought, the concept of structural equivalence \parencite{lorrain_structural_1971}, 
blockmodelling \parencite{white_social_1976}, and its generative application in stochastic blockmodels \parencite{nowicki_estimation_2001} can be applied to the missingness indicator. The resulting model gives us a framework to assert differences in the probability of a missing tie variable depending on their block membership. For example, this lets us assume that missing tie variables are more likely between blocks than within them. Clustering methods or other community detection algorithms can be used to identify structurally similar nodes, and in the case of some blockmodels, structurally equivalent nodes. 

The stochastic blockmodel may be applicable to simulate missingness with block structure. This requires the researcher to know roles and positions of certain actors in the covert network, which is ideally informed with a hypothesis to quantify how likely missing tie variables are for specific roles relative to other roles. Stochastic blockmodels can be used to identify latent classes for the nodes and the heterogeneity in the degree of the missing tie variables can be independent conditional on the node class. Statistical approaches to blockmodelling can also be understood in the context of methods that describe latent properties of the missingness, or more generally social space methods.

Social or latent space models generally rely on an approximation of a latent space of the missingness and can weigh the latent vectors of the missingness to affect the probability of a missing tie variable through a function $h(.,.)$. Adapting from Sosa \& Buitrago (2020), independently for each $d_{ij}$, 

$$\Pr(d_{ij} = 1 | c_{ij}, \alpha, \beta, \gamma_{ij}) = logit^{-1}(\alpha + \beta^{\top}c_{ij} + \gamma_{ij}),$$
$$\gamma_{ij} = h(u_i, u_j).$$

\noindent where $\mathbf{\beta} = (\beta_1, ..., \beta_P)$ is a vector of fixed effects and $c_{ij}$ refers to the covariates for each tie variable. The parameter $\gamma_{ij}$ is a random effect representing any patterns in the data unrelated to the predictors in the model including functions of the latent space. If the random effects $\gamma_{ij}$ is jointly exchangeable for any permutation of $i$ and $j$, then $h(.,.)$ is a symmetric function such that $\gamma_{ij}$ = $h(u_i, u_j)$, where $u_i$ consists of s sequence $u_{i1}, ..., u_{il}$ of $L$ independent latent random variables defining the latent space which is defined in $\mathbb{R}^L$. The impact of the latent variables on the overall model as defined above depends on the form of $h(.,.)$. 

It is through $h(.,.)$ that we can reflect how missingness may be heterogeneous in the data.  Other approaches have described $h(.,.)$ as a function defining the distance or angle between any two nodes in the latent space \parencite{hoff_latent_2002}. A difficulty when dealing with latent space models, or latent variables in general, is the difficulty of interpreting the results. Parameter values of the model would be able to tell us probabilities of configurations of missing tie variables (and consequently, expected counts) given the latent space, but it may not be clear what exactly this space represents. It is however, a statistical model that conditionally accounts for the dependence in the network through the latent space while maintaining convenient independence properties.

When we start choosing which missingness model would fit a given scenario, we can start from what each of the missingness models imply about the missingness mechanisms over and above the missingness assumption. The independent or Bernoulli case implies that the missingness is completely at random, and more importantly implies that missingness occurs independently for each tie variable. Heterogeneity can be introduced here using heterogeneous Bernoulli models, thus letting the missingness to be dependent on either itself or external covariates. 

The latent model is a fairly complicated case as the model is conditionally independent. As the nodes’ latent space positions are used to explain the occurrence of ties through some probability function of the latent space positions between two nodes, and, by definition, a latent space is an abstract unobserved algebraic space, this leads to some complicated implications when we use them for a missingness model. 

We are assuming that there is indeed a latent space (or a combination of some unobserved dimensions) that can explain why some nodes have more easily observable tie variables with specific nodes through some function of the two nodes’ positions in this latent space. We are implicitly making the assumption that these dimensions are informative or reflective of some empirical phenomena if we wanted to explain their effects. If we wanted to simply have them be abstract objects for the utility of having an abstract space (e.g., prediction purposes that use the latent space), we may be able to use the latent model. Nonetheless, this latent space reflects an abstract mathematical representation of sampling biases when networks are constructed. As the latent space is mathematically defined, the missingness of a particular tie variable would be dependent on two mathematical objects, the missingness models’ model parameters ($\alpha, \beta$), and the latent model’s specifications ($h(.,.)$). 

\subsection{Relaxing independence}

The models described in the section above all assume that the probability that a tie variable is missing is completely independent of the missingness of other tie variables. For example, a high $\beta_i$ would suggest that tie variables $d_{ij}$ and $d_{ik}$ were more likely to be missing as a function of the $\beta_i$ value without acknowledging that $d_{ij}$ and $d_{ik}$ share a node $i$. If we were to consider this assumption when taking into account the construction of a covert network, we would need to be confident that all the possible sources of missingness were dyadic in nature. That is, that all the missingness is affected by single tie varaibles and not adjacent ties simultaneously. This would mean that any process that biases the observability of the data affects (the observation of) tie variables independently of each other. When we consider that social network data describe dependencies between individuals, the independence assumption is difficult to justify for both the tie variables and missingness indicators. For example, it would be odd if the tie variables $x_{ij}$ and $x_{jk}$ were observed but $x_{ik}$ were missing (i.e., $d_{ik} = 1$). Assuming the missingness of $d_{ik}$ to be independent of the other relationships of actor $i$ or $k$ would be a very difficult assumption to support, especially when tie variables involving the same actors were observed, $d_{ij} = 0$ and $d_{jk} = 0$.

If we consider the possibility that missingness of tie variables can depend on the missingness of other tie variables, we are able to propose more flexible assumptions for the processes by which missingness might occur. Missingness may in part be the result of investigatory (or technological or judicial) processes that produce gaps in the observation of a network that involve more than one dyadic tie at a time. Therefore, just as a statistical model for  \textbf{X} will test for various social processes affect tie formation or maintenance, so does a statistical model for \textbf{D} account for various observational processes that produce missingness in a network. For example, just as popular individuals are surrounded by ties, so might evasive individuals be surrounded by missing observations.

More directly, if a covert actor is \emph{successful} in keeping their interaction with another actor hidden, then they are more likely to be successful in keeping their interaction with yet further actors hidden. We are not specific about the extent to which missingness stems from actors actively hiding or the data collector(s) not paying equal attention to all dyads. We are simply saying that the resulting missingness is the result of an information gap in the knowledge about the actor in one dyad itself, which can propagate to the other dyads this actor is involved in.  The examples here are just a few of the many cases where missingness is unlikely to be dyadically independent.

To describe dependence between missing tie variables, we propose that Markov dependency assumptions usually made for tie variables can be extended to missing data indicators. The Markov dependency assumption states that two tie-variables are conditionally independent unless they share a node, given everything else \parencite{frank_markov_1986}. By extending this assumption to missing data indicators, missing tie variables that share the same node are conditionally dependent on each other, conditional on the rest of the missing data indicator $\mathbf{D}$. To unpack this assumption, we may suggest if actor $a$ had a missing tie variable to actor $b$, it may be more likely that actor $a$ would also have a missing tie variable to another actor $c$. 

A potential setting where the dependency of missingness on other missing could be attentional biases (i.e., a 'spotlight' effect) in which observation is skewed towards certain nodes or potential ties, and away from others. A keen reader may notice that the bias mentioned describes an observational mechanism and not explicitly a missingness mechanism. As represented in the definition of the missingness indicators $d_{ij}$, we note that the two concepts are fundamentally related. If we considered that these observations often occur under the constraint of limited resources, as certain observations are more likely to be observed, other observations are less likely to be observed and thus more likely to be missing.

Another practical avenue to justify the Markov dependency assumptions for missing data indicators are sunken cost fallacies or the constraint of limited resources. These may be reflecting generalised observational biases, but constrained with limited resources (i.e., every possible tie variable cannot be evaluated as it would be costly). An investigator might have their own intuitions to narrow down who they perceive to be important or central members of a covert network. As a consequence, they may focus on their potential suspects and pay less attention to other members. Consequently, the missing tie variables proliferate as evaluating a seemingly-irrelevant relationship requires expenditure of limited resources, making it less likely for said relationship to be observed and thus suggesting that any other relationship involving these two individuals is more likely to be missing.

Alternatively, there may be biases towards surveilling certain nodes, but also certain social settings in which small groups form. This may result in biases in the observation of  clusters of individuals.

In the scenario above, we describe two possible cases where subgroups can be either secretive or salient. To represent either case, we further elaborate the basic Markov assumption and illustrate the occurrence of triadic structures in the missingness mechanism to capture a 'cluster' of missing tie variables. On a more theoretical level, we may consider the context or setting in which the subgroup is formed. If we consider nodes $i, j,$ and $k$ to belong to a more secretive subgroup who only meet to discuss clandestine covert matters, an analyst may miss node $i$ and its ties, $d_{ij}d_{ik} = 1$. As a consequence of missing node $i$ and its partners $j$ and $k$, the partners' tie is likely to be missing as well $d_{jk} = 1$. Rebinarising the missingness mechanism to a sampling mechanism can evaluate the sampling of salient subgroups.

To accommodate departures from independence of missingness indicators, we propose to model $\mathbf{D}$ using the Exponential Random Graph Model (ERGM)

\begin{equation}
    \Pr(\mathbf{D} = d, \psi) = \exp\{\psi^{\top}z(d) - \kappa(\psi)\}.
    \label{equation:ergm}
\end{equation}

\noindent The function $z()$ describes the model specification for the missingness model and $\kappa$ is the normalising constant $\log \sum_{D \in \mathcal{D}} \exp\{\psi^{\top}z(D)\}$. While the ERGM is typically used as a statistical model to describe network structures as a product of locally emergent social processes implied by $z()$, when we apply it to the missingness indicator matrix, ERGMs allow for a wide variety of dependence assumptions. We note that the interactions specified in $z()$ and included in the ERGM should follow the statistical principle of hierarchy (e.g., if three-stars are included, so should two-stars), but the difficulty of simulating strictly Markov assumptions while avoiding phase transitions is a known issue \parencite{handcock_assessing_2003}. We follow standard practice and use geometrically weighted (or alternating) statistics 
 that are commensurate with Markov dependence but also more stable \parencite{snijders_new_2006}. Not only can the ERGM include a variety of dependence structures \parencite{pattison_9_2002, lusher_modeling_2012} in the sampling and missingness mechanism, a variety of covariates for dyads and nodes can also be included. These can be attributes of the nodes or a relevant dyadic covariate.

While the method is statistically flexible, the most important consideration is the context of the missingness. There is no single rule for the applicability of any missingness assumption and the application of particular missingness assumptions is subject to further research on policing. If the context suggests an independent mechanism, such as data randomly entered incorrectly, then an independent missingness model would apply. If the context suggests a strong tendency for missing tie variables to cluster with other missing tie variables then the missingness model should reflect this tendency. The choice of which model would be the most appropriate is best evaluated with as much information about the missingness generating mechanism as possible. Practically speaking this may be qualitative information (e.g., informant reports) of yet-unobserved tie variables or acknowledgement of a biased sampling mechanism.  

\section{Conditioning on the network}

To assume $\mathbf{D}$ to be completely independent of the true network $\mathbf{X}$ is an unrealistic assumption. From a generative model standpoint, by using information about the true network, we are able to describe how certain properties of the true network affect the missingness of tie variables. We have previously suggested statistical models that generate missing tie variables through statistical parameters or through some covariate information. There are multiple ways for the true network to be accounted for in the missingness model. By using the true network as a dyadic covariate, we can reflect empirical biases involving the true network as a statistical model. An example is a bias for missing tie variables to be edges than null ties. This model can be expressed as

$$Pr(d_{ij} = 1| \mathbf{X} = x_{ij}) = {\mathrm{logit}}^{-1}(\alpha + \beta_1x_{ij}), $$

\noindent applied independently for all tie variables $d_{ij}$. A positive $\beta_1$ would be mean that edges are more likely to be missing than null ties. Conveniently, this case can also be understood as an entrainment effect from a dyadic covariate in an ERGM framework. Beyond a dyadic covariate, we can use functions of the network as covariates for the missingness model. An example of this case could be that nodes with higher degree are more likely to have missing ties. This case can be expressed as

$$Pr(d_{ij} = 1|\mathbf{X} = x_{ij}) = {\mathrm{logit}}^{-1}[\alpha + \beta_1x_{ij} + \beta_2(x_{i+} + x_{+j})],$$

\noindent applied independently for all tie variables $d_{ij}$. In this case, a positive $\beta_2$ would suggest that nodes with higher degree would be more likely to have missing tie variables. Here the network reduces to a monadic effect similar to the node covariates in the independent models (all else being equal; also applies to the $\beta$ model). An example of another function of the network, could be to use the distance in $\mathbf{X}$ of the dyad $\{i,j\}$, from some focal actor in the network \parencite[as in ][, where the distance from Noordin Top is assumed to decrease visibility]{koskinen_bayesian_2019}.

One last example would be the case where the probabilities $d_{ij}$ both depend on the network $\mathbf{X}$ and other missing tie variables. In this case, we can no longer express missing tie variable probabilities $p_{ij}$ independently for each tie variable and instead would need to model the entire matrix $\mathbf{D}$, which in exponential family form would be

$$P(\mathbf{D} | \mathbf{X}, \theta) = exp\{\theta^{\top}z(\mathbf{D}) + {\theta_a}^{\top}f(\mathbf{D}, \mathbf{X}) - \kappa(\theta)\},$$

\noindent where $z(\mathbf{D})$ is only a function of $\mathbf{D}$, but $f(\mathbf{D},\mathbf{X})$ is simultaneously a function both of the network $\mathbf{X}$ as well as the missing indicators $\mathbf{D}$. In this last case, we can easily distinguish between endogenous and exogenous effects in the missingness mechanism. In the missingness model, we defined endogenous effects to be cases where the missing tie variables depend on other missing tie variables, e.g. through interactions $d_{ij}d_{ik}$. In contrast, exogenous effects occur when the missing tie variables depend on external variables, which in the case above is the network tie variable, and interactions of the type $d_{ij}x_{ij}$. The exogenous effects here can encompass multiple data structures\footnote{We use the term \emph{exogenous} for clarity, but the function  $f(\mathbf{D},\mathbf{X})$ could of course also include interactions of type $d_{ij}d_{ik}x_{jk}$}. This can be the network $\mathbf{X}$ which lets us evaluate the probability of missing tie variables being ties in the network ($d_{ij}x_{ij})$. It also lets us include dyadic $\mathbf{C}=(C_{ij})$ or monadic actor $\mathbf{c}=(C_{i})$ covariates and how these can affect missing tie variables.

\subsection{General case of the missingness model}

When we consider the different data structures that the missingness model can be conditioned on, we can describe a general case of the missingness model for network data. This can be described as
\begin{equation}
    P(\mathbf{D} | \mathbf{X}, \beta, \psi, \theta) =  \exp(  \psi^{\top} f(\mathbf{D}) +\beta^{\top} f(\mathbf{D}, \mathbf{C})+ \theta^{\top} f(\mathbf{D}, \mathbf{X}) - \kappa( \beta, \psi, \theta)).
    \label{equation:missGeneral}
\end{equation}

\noindent In Equation \ref{equation:missGeneral}, the parameters $\psi$ refer to endogenous effects where the missingness mechanism is dependent on itself. The parameters $\beta$ refer to interactions between the missingness mechanism and dyadic covariates $\mathbf{C}$. The parameters $\theta$ refer to interactions between the missingness mechanism and the true network. Lastly, $\kappa$ refers to the normalising constant of the specified model, $\kappa( \beta, \psi, \theta) = \log \sum_{\mathbf{D} \in \mathcal{D}} \exp(  \psi^{\top} f(\mathbf{D}) +\beta^{\top} f(\mathbf{D}, \mathbf{C})+ \theta^{\top} f(\mathbf{D}, \mathbf{X}))$.

\subsection{Missingness assumptions and the network}

The missingness mechanism may or may not be conditional on the network. We follow the classic Rubin (1976) definitions of missingness.
Firstly, when the missingness mechanism is completely independent to the network, data are said to be missing completely at random (MCAR),

$$P(\mathbf{D}|\mathbf{X}, \psi) = P(\mathbf{D}|\psi). $$

\noindent The MCAR assumption applies to any endogenous dependence assumptions the missingness mechanism has with itself as these processes do not depend on the network. 

If the missingness mechanism is only dependent on the observed parts of the network, $\mathbf{X}_{obs}=(X_{ij}:d_{ij}=0)$, 
\begin{equation}
   P(\mathbf{D}|\mathbf{X}, \psi) = P(\mathbf{D}|\mathbf{X}_{obs}, \psi),
    \label{equation:missMAR}
\end{equation}
we say that data are missing at random (MAR). Lastly, if the missingness mechanism could not be simplified and were dependent on both observed and missing parts, $\mathbf{X}_{mis}=(X_{ij}:d_{ij}=1)$, of the network,
$$P(\mathbf{D}|\mathbf{X}, \psi) = P(\mathbf{D}|\mathbf{X}, \psi). $$
data are missing \emph{not} at random (MNAR).
\noindent Generally speaking, it would be safer to assume MNAR missingness as the MAR assumption is far less of a general assumption. 

We can think of the ERGM as the most flexible model for our purposes in multiple ways. The ERGM (Eq.~\ref{equation:missGeneral}) incorporates the homogeneous and heterogeneous Bernoulli models, as well as the Markov model and the model that conditions on the true network as its special cases. We can also think of the ERGM as the most flexible model in its capacity of being the maximum entropy model \parencite{banks_metric_1998} that is able to parameterise departures from homogeneity and independence. Furthermore, the model can also be used to parameterise the extent to which the missingness model violates conventional missingness assumptions of M(C)AR. How Equation (\ref{equation:missGeneral}) generalises the conventional missingness assumptions is summarised in Table \ref{table:missAssumps}. Assuming an ERGM for $\mathbf{D}$, can thus be seen as giving missing data maximal freedom (entropy), subject to the constraints of the stipulated dependence assumptions.

\begin{table}[H]
\centering
\begin{tabular}{ll}
\hline
Parameter values & Missingness assumption \\  \hline
$\theta = 0$                & MCAR                   \\
$\theta = 0, \beta = 0 $               & Heterogeneous MCAR      \\
$\theta = 0, \beta = 0, \psi = 0 $               & Homogeneous MCAR   \\
$\theta = 0, \beta \neq 0$            & MAR                    \\
$\theta \neq 0$                & MNAR                   \\

\hline
\end{tabular}
\caption{Table describing parameter values in Equation \ref{equation:missGeneral} and their corresponding missingness assumption}
\label{table:missAssumps}
\end{table}

The MAR assumption (Equation \ref{equation:missMAR}) greatly relaxes the MCAR assumption, yet inference for ERGM (for $\mathbf{X}$) only requires a slight modification of the likelihood equations \parencite{handcock_model-based_2007} and is implemented in \texttt{statnet} \parencite{handcock_ergm_2018}. While convenient and seemingly general, the only extant example of a MAR mechanism for networks is snowball sampling and other link-tracing designs \parencite{thompson_model-based_2000}.

To illustrate what the MAR assumption implies, we adapt a bivariate example of MAR for non-network data in \textcite{little_statistical_1987} to network data. The following example is an attempt at demonstrating a MAR mechanism for a pair of two binary tie variables $x_{ij}$ and $x_{ik}$. As the two tie variables share a node, the MAR example below can also be interpreted as a specific case of Markov dependence where the missing tie variables depend its adjacent observed tie variable. Let us then consider a model for a pair of tie variables,

$$\Pr(d_{ij} = r, d_{ik} = s | x_{ij}, x_{ik}, \psi) = g_{rs}(x_{ij}, x_{ik}, \psi), r,s \in \{0, 1\},$$

\noindent where the model independently describes missingness for each pair of tie variables. Due to the MAR assumption, a missing tie variable cannot depend on itself as it is not observed. Instead, the missing tie variable depends on its paired observed tie variable. We first address the example when either $d_{ij} = 1$ or $d_{ik} = 1$. In the first case of $d_{ij} = 1$, the conditional probability of $d_{ij}$ being missing is only dependent on the observed $x_{ik}$ and the parameters $\psi$,

$$g_{10}(x_{ij}, x_{ik}, \psi) = g_{10}(x_{ik}, \psi).$$

\noindent The model $g_{10}$ removes $x_{ij}$ from its arguments as $d_{ij}$ cannot depend on $x_{ij}$ due to the tie variable being missing and the MAR assumption. For the other case of $d_{ik} = 1$, the same logic applies for the other tie variable,

$$g_{01}(x_{ij}, x_{ik}, \psi) = g_{01}(x_{ij}, \psi).$$

\noindent While the examples here address pairs of tie variables, ultimately the same assumption is made where the missingness of the missing tie variable depends on its paired observed tie variable. 

If both tie variables are missing, by definition of MAR the probabilities $d_{ij} = 1$ and $d_{ik} = 1$ cannot depend on either $x_{ij}$ of $x_{ik}$ due to the tie variables being missing and we are left with $\psi$,

$$g_{11}(x_{ij}, x_{ik}, \psi) = g_{11}(\psi).$$

\noindent Using the rule of complements after defining the other three probabilities, the probability that both $x_{ij}$ and $x_{ik}$ are observed is given by

$$g_{00}(x_{ij}, x_{ik}, \psi) = 1 - g_{10}(x_{ik}, \psi) - g_{01}(x_{ij}, \psi) - g_{11}(\psi). $$

\noindent While the above is clearly an example of MAR and not MCAR, this requires a strict and unrealistic assumption for the way pairs of tie variables are observed. It is further clear that to construct MAR, you have to relax independence. While this interdependence has a flavour of Markov dependence, it is not sufficient as the dependence is strictly dyadic and not easily extended to extra-dyadic dependencies.
We discuss a few more cases of MAR with binary variables in Appendix A. 

If we had node or dyadic covariates, and if we assumed these covariates explained the missingness, we can rely on the covariates for the MAR assumption. If we revisited the MAR example above independently for tie variable $x_{ij}$ and corresponding covariate $c_{ij}$,

$$Pr(d_1 = r, d_2 = s | x_{ij}, c_{ij}, \beta) = g_{rs}(x_{ij}, c_{ij}, \beta), r,s \in \{0, 1\},$$
$$g_{10}(x_{ij}, c_{ij}, \beta) = g_{10}(c_{ij}, \beta),$$
$$g_{01}(x_{ij}, c_{ij}, \beta) = g_{01}(x_{ij}, \beta),$$
$$g_{11}(x_{ij}, c_{ij}, \beta) = g_{11}(\beta),$$
$$g_{00}(x_{ij}, c_{ij}, \beta) = 1 - g_{10}(c_{ij}, \beta) - g_{01}(x_{ij}, \psi) - g_{11}(\beta). $$

\noindent The example then presents a much more straightforward picture where the value of the covariate $c_{ij}$ can account for the probability of its corresponding tie variable $x_{ij}$ being missing and vice versa. An example of this could be members of a covert network with certain roles may be less likely to have their ties observed. An active drug courier may be easier to observe in contrast to the more secretive suppliers.

One other example of MAR in network sampling as described in \textcite{handcock_modeling_2010} suggests that multi-wave link tracing designs such as snowball sampling can be mathematically formulated in a way to be satisfied under MAR. However, a required assumption for the sampling design to only be dependent on the observed data is the complete observation of ties for sampled nodes. This may be a difficult assumption when considering the network construction process for covert networks.

In summary, despite MAR being the conventional and convenient assumption, 
empirically motivating a mechanism that follows MAR can be difficult. Having additional information, such as node or dyadic covariates, could be helpful when assuming MAR. We also note that heterogeneity through endogenous dependence assumptions in the missingness indicator can be accounted for under MCAR. Generally speaking, MNAR would be the safest assumption to assume if the missingness were dependent on the network. However, practically speaking it can be difficult to assess as the MNAR process is inherently unknown. 

\subsection{Estimating ERGMs with missing values}

When it comes to readily available software that estimates network statistical models while acknowledging the missingness, we revisit the 'ergm' package in R. The routines within the ergm package relies on likelihood-based inference \parencite{krivitsky_likelihood-based_2023} for the estimation of ERGMs, which is based on the model-based framework developed in \parencite{handcock_modeling_2010}. To clarify the likelihood-based framework for ERGM estimation, when data are fully observed, inference for model parameters $\eta$ and with statistics $s()$ can be based on the fully observed likelihood

$$L(\eta ;\mathbf{X}) \propto P(\mathbf{X}| \eta)
= \exp\{ \eta^{\top} s(\mathbf{X}) - \kappa(\eta) \}.$$

\noindent However, when data are partially observed, we follow the likelihood-based inference for incomplete data as described in \textcite{little_statistical_1987}. The inferential framework states that the missingness mechanism is \textit{ignorable} when the missing data are MAR and when the parameters for the data and missingness are distinct so that the joint parameter space is $\boldsymbol{\eta} \times \Psi$. We can thus define the \textit{face-value likelihood} for inference purposes:

$$L(\eta;\mathbf{X}_{obs}) \propto \sum_{\mathbf{X}_{mis}} P(\mathbf{D}| \mathbf{X}_{obs}, \mathbf{X}_{mis}, \psi) P(\mathbf{X}_{obs}, \mathbf{X}_{mis}| \eta). $$

\noindent The face-value likelihood marginalises over all the missing tie variables, thus only relying on the observed tie variables to model the model parameters. By assuming the missing mechanism to be ignorable for inference purposes, we factor out the model for the missingness as it turns into a constant, 

\begin{equation}
\begin{split}
L(\eta;\mathbf{X}_{obs}) & = P(\mathbf{D}|\mathbf{X}_{obs}, \psi) \sum_{\mathbf{X}_{mis}} P(\mathbf{X}_{obs}, \mathbf{X}_{mis}| \eta) \\
                            & \propto \sum_{\mathbf{X}_{mis}} P(\mathbf{X}_{obs}, \mathbf{X}_{mis}| \eta).
\end{split}
\label{equation:faceLike}
\end{equation}
Equation \ref{equation:faceLike} results in the likelihood equation
\begin{equation}
E_{\eta}\left[s(\mathbf{Y})\right]=E_{\eta}\left[s(\mathbf{Y}) \mid \mathbf{X}_{obs}\right],
\label{equation:MLE}
\end{equation}
where $s()$ is the vector of statistics of the data model and  $\mathbf{Y}$ is defined by the model $P(\mathbf{Y}|\eta)$.
\noindent We can see that Equation \ref{equation:faceLike} is only possible when we make the MAR assumption, and thus we can only use (Equation \ref{equation:MLE}) under MAR. In other words, the formulation above ignores any information about $\eta$ available in $\mathbf{D}$. At the time of writing, there are no readily available software to circumvent making the ignorability assumption. This further supports the value of model-based representations to evaluate missingness mechanisms that are not ignorable.

\section{Simulation study}

To demonstrate how missingness can be expressed and demonstrate their effects on covert network data, we assumed different missingness models and degraded empirical networks with them. These simulations follow previous Monte Carlo studies that have evaluated the effects of missing data in network analysis \parencite{kossinets_effects_2006, huisman_imputation_2009}. This study specifically replicates the scenario of estimating a model assuming either MAR or using zero imputation under different actual conditions of missingness. Ideally, the statistical assumptions made through the specification of the missingness model reflect realistic observational biases encountered in the network construction process, thus representing the biased and layered construction process as described above. Furthermore, simulating the missingness offers precision in the types of biases that are being assumed in the construction of the network which allows us to systematically investigate the consequences of various biases on constructed covert networks. We investigated four different elements when specifying the missingness. These four elements are (a) the estimation model, (b), the missingness model, (c) the representation of missingness (MAR or zero imputation), and (d) the proportion of missingness. We assume that the empirical covert networks are the true network. We make this assumption to have a reference point to let us evaluate the consequences of any missingness encountered.

\subsection{Empirical networks}

\begin{figure}[H]
    \centering
    \includegraphics[width=0.8\linewidth]{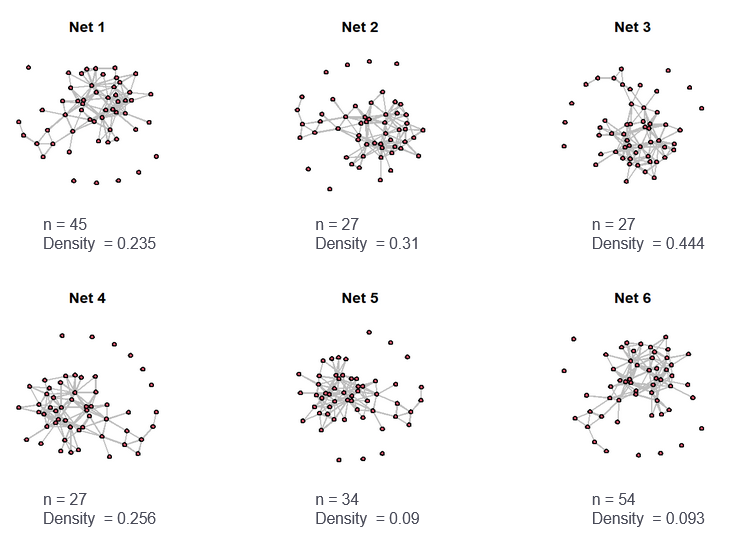}
    \caption{The six empirical covert networks used in the simulation studies with node sizes and densities}
    \label{fig:netsUsed}
\end{figure}

Six empirical covert networks were evaluated in this study. They  were all obtained from the UCINet database of Covert networks \parencite{borgatti_ucinet_2002}. These networks were chosen as they are all unimodal and cross-sectional networks that did not have too many nodes with some collected attribute data. The specific networks used as seen in Figure \ref{fig:netsUsed} are:
\begin{enumerate}
    \item Jemaah Islamiyah bombing on Christmas Eve 2000 
    \item Jemaah Islamiyah bombing on Bali 2002 
    \item Jemaah Islamiyah bombing on the Australian Embassy in 2004 
    \item Jemaah Islamiyah bombing on Bali 2005 
    \item Hamburg cell associated with sleeper Al Qaeda cell around 9/11 bombings in 2001 
    \item Co-offending London gangs between 2005 and 2009

\end{enumerate}

There are four steps to the simulation study. The first step was to estimate models on the six empirical networks chosen. This was to serve as a reference point of the model for the network, i.e. the inference an analyst would make, had they had incomplete information. Two different estimation models were estimated for the six networks using the 'ergm' package in R \parencite{handcock_ergm_2018}. We chose the ERGM as the baseline for comparison as it is considered to be the gold standard of network statistical models \parencite{lusher_exponential_2013}. As described above, when estimating ERGMs with missing values, the package follows \textcite{handcock_modeling_2010} in its use of the face-value likelihood, solving the likelihood equation. Note that \textcite{handcock_modeling_2010} rely on the \textcite{geyer_constrained_1992} MCMC MLE. Networks 1 to 5 were estimated using a simpler purely structural model while Network 6 was estimated with covariates. The results of the 'true' (or 'gold standard') model can be seen in Table \ref{table:trueModel}. 

\begin{table}[H]
\centering
\begin{tabular}{llcccccc}
\hline
Parameter  & Statistic          & Net 1                                                  & Net 2                                                  & Net 3                                                   & Net 4                                                  & Net 5                                                  & Net 6                                                  \\ \hline
$\eta_1$ & Edges                 & \begin{tabular}[c]{@{}c@{}}-2.93\\ (1.89)\end{tabular} & \begin{tabular}[c]{@{}c@{}}-1.75\\ (2.61)\end{tabular} & \begin{tabular}[c]{@{}c@{}}-14.61\\ (3.40)\end{tabular} & \begin{tabular}[c]{@{}c@{}}-0.77\\ (1.78)\end{tabular} & \begin{tabular}[c]{@{}c@{}}-4.06\\ (0.37)\end{tabular} & \begin{tabular}[c]{@{}c@{}}-6.31\\ (1.02)\end{tabular} \\
$\eta_2$ & Altkstar(2)           & \begin{tabular}[c]{@{}c@{}}-2.16\\ (0.48)\end{tabular} & \begin{tabular}[c]{@{}c@{}}-2.10\\ (0.64)\end{tabular} & \begin{tabular}[c]{@{}c@{}}-1.72\\ (0.80)\end{tabular}  & \begin{tabular}[c]{@{}c@{}}-1.88\\ (0.44)\end{tabular} & \begin{tabular}[c]{@{}c@{}}-0.14\\ (0.23)\end{tabular} & \begin{tabular}[c]{@{}c@{}}-0.51\\ (0.23)\end{tabular} \\
$\eta_3$ & Gwesp(log(2))         & \begin{tabular}[c]{@{}c@{}}4.70\\ (0.41)\end{tabular}  & \begin{tabular}[c]{@{}c@{}}4.05\\ (0.48)\end{tabular}  & \begin{tabular}[c]{@{}c@{}}9.70\\ (1.38)\end{tabular}   & \begin{tabular}[c]{@{}c@{}}3.10\\ (0.37)\end{tabular}  & \begin{tabular}[c]{@{}c@{}}1.49\\ (0.33)\end{tabular}  & \begin{tabular}[c]{@{}c@{}}1.29\\ (0.18)\end{tabular}  \\
& nodecov(Age)          &                                                        &                                                        &                                                         &                                                        &                                                        & \begin{tabular}[c]{@{}c@{}}0.08\\ (0.02)\end{tabular}  \\
& absdiff(Age)          &                                                        &                                                        &                                                         &                                                        &                                                        & \begin{tabular}[c]{@{}c@{}}-0.18\\ (0.04)\end{tabular} \\
& nodecov(Arrests)      &                                                        &                                                        &                                                         &                                                        &                                                        & \begin{tabular}[c]{@{}c@{}}0.07\\ (0.02)\end{tabular}  \\
& absdiff(Arrests)      &                                                        &                                                        &                                                         &                                                        &                                                        & \begin{tabular}[c]{@{}c@{}}-0.01\\ (0.02)\end{tabular} \\
& nodecov(Convictions)  &                                                        &                                                        &                                                         &                                                        &                                                        & \begin{tabular}[c]{@{}c@{}}-0.08\\ (0.03)\end{tabular} \\
& absdiff(Convictions)  &                                                        &                                                        &                                                         &                                                        &                                                        & \begin{tabular}[c]{@{}c@{}}-0.07\\ (0.04)\end{tabular} \\
& nodematch(Prison)     &                                                        &                                                        &                                                         &                                                        &                                                        & \begin{tabular}[c]{@{}c@{}}0.11\\ (0.19)\end{tabular}  \\
& nodematch(Birthplace) &                                                        &                                                        &                                                         &                                                        &                                                        & \begin{tabular}[c]{@{}c@{}}0.84\\ (0.15)\end{tabular}  \\ \hline
\end{tabular}
\caption{Table of model estimates for complete network to be used as the baseline $\eta$}
\label{table:trueModel}

\end{table}

After the 'true' models of the data were estimated, we proceeded to simulate a variety of missingness models. We estimated four different missingness models of increasing complexity, a completely independent model, a latent space model to induce heterogeneity while remaining conditionally independent, an ERGM with endogenous dependence assumptions that follows MCAR, and an ERGM conditional on the network that follows MNAR. The latent space model included a simulated latent space for all nodes in each network. The latent model was also specified to suggest that nodes with greater Euclidean distance from each other in the latent space are more likely to have missing tie variables. 50 realisations of each model were taken to capture the variance in the realised missingness indicators. Three different amounts of missingness were also taken to evaluate the effects of varying quantities of missingness. These proportions are small (10\%), medium (35\%), and large (60\%). 

The realisations of the missing data indicators $\mathbf{D}$, were used to degrade the true network, and missingness was represented in two ways. Firstly, missingness was (properly)  represented as missing (i.e., 'NA', as in Eq.~\ref{def:xobsmiss}) and keeping the tie variable missing, something which we refer to as 'Miss'. Second, for comparison, we employ the common practice of \emph{zero imputation}, where missing tie-variables are set to 0. We refer to this implied assumption as 'Zero'. After the networks were degraded, we then re-estimated the same models for the degraded networks to gauge the effects of the missingness model on the model parameters. For the 'Zero' condition, we estimate the ERGM for data `as is', i.e. assuming data are the observed ties and everything else 0. For the 'Miss' condition, however, we estimate the ERGM under the assumption of MAR, using the likelihood (Equation \ref{equation:faceLike}), for \emph{all} missing data generating mechanisms. The last step of the simulation was to analyse the re-estimated models. Further details of the missingness model specification can be found in Table \ref{table:ergmMissParams}.

\begin{table}[H]
\begin{tabular}{llll}
\hline
Parameters & Statistics       & \multicolumn{2}{c}{Missingness model}          \\ \hline
               &  & M(C)AR ERGM & MNAR ERGM \\
$\psi_1$ & Edges   $\sum_{i < j} d_{ij}$         & 0         & 0                      \\
$\psi_2$ & GWDegree   $e^{\alpha}\sum_{k = 1}^{n-1} \{1 - (1 - e^{-\alpha})^k\}D_k(d)$      & 2         & 0.4             \\
$\psi_3$ & GWESP    $e^{\alpha} \sum_{i = 1}^{n-2} \{1 - (1 - e^{-\alpha})^i \} sp_i$        & 2         & 0.5               \\
$\theta_1$ & Entrainment   $\sum_{i < j} d_{ij} x_{ij}$     &           & 0.8              \\
$\theta_2$ &Degree covariate $\sum_{i < j} d_{ij} \sum_i x_{ij}$ &           & 0.2                \\ \hline
\end{tabular}
\caption{Table of ERGM parameters for the missingness models}
\label{table:ergmMissParams}

\end{table}

According to the ergm R package manual, we expect the introduction of any amount of missingness will make the estimation of the model more difficult as the routines in place to handle missing data may introduce complicated steps in the MCMLE, as you need to not only sample from $P(\mathbf{X}\mid\eta)$ for the left-hand side of (Equation \ref{equation:MLE}), but also sample from $P(\mathbf{X}_{mis} \mid \mathbf{X}_{obs},\eta)$ for the right-hand side of (Equation \ref{equation:MLE}) \parencite{handcock_ergm_2023}. We also expect density-based measures to be particularly affected when the missing values are set to 0.

\section{Implementation and examples}

In total, 50 realisations of 4 different missingness models, with 3 proportions of missingness and 2 different representations of missingness, resulting in a total of 1200 different realised missingness indicators were used to degrade each of the 6 empirical networks. 

There were a total of 7200 degraded networks, and we summarise the key insights from various plots and inspections. The complete results and code are available on request from the authors.

\subsection{Failure rates}

ERGMs are known to be difficult to estimate and some networks may be degenerate \parencite{handcock_assessing_2003}. As we are following the scenario of an analyst using standard available software in the estimation of ERGMs, we calculated a 'failure rate' value to indicate the propensity for the model estimation to fail. The computation of the failure rate is to illustrate the difficulties encountered when estimating ERGMs with partially observed data using standard software. This value was calculated by tabulating the number of successful re-estimations and dividing it by the total amount of estimations performed for a given degraded network. A model re-estimation was considered to be a failure when the statnet algorithm fails to compute and exits with one of the errors as below. These errors include (a) the inability of the MCMC sampling to mix resulting in the inability to reach the desired effective same size was not able to be reached for the specified number of iterations, (b) the MCMLE estimation getting stuck due to excessive correlation between model terms, or (c) Matrix a has negative elements in the diagonal resulting in NaNs produced.

We note in particular the error (c) is a common and known issue with the use of the face-value likelihood. The Hessian of the face-value likelihood is

\[
Cov\left[s(\mathbf{Y}),s(\mathbf{Y}) \mid \mathbf{X}_{obs}\right]-Cov\left[s(\mathbf{Y}),s(\mathbf{Y})\right].
\]

\noindent We can see that the the when the conditional expectation numerically evaluates to something greater than the corresponding elements of the unconditional expectation, the negative Hessian is not positive definite. This pragmatically amounts to additional numerical computations so the negative Hessian is positive definite and the algorithm can proceed to estimate the model.

\begin{figure}[H]
    \centering
    \includegraphics[width=10cm]{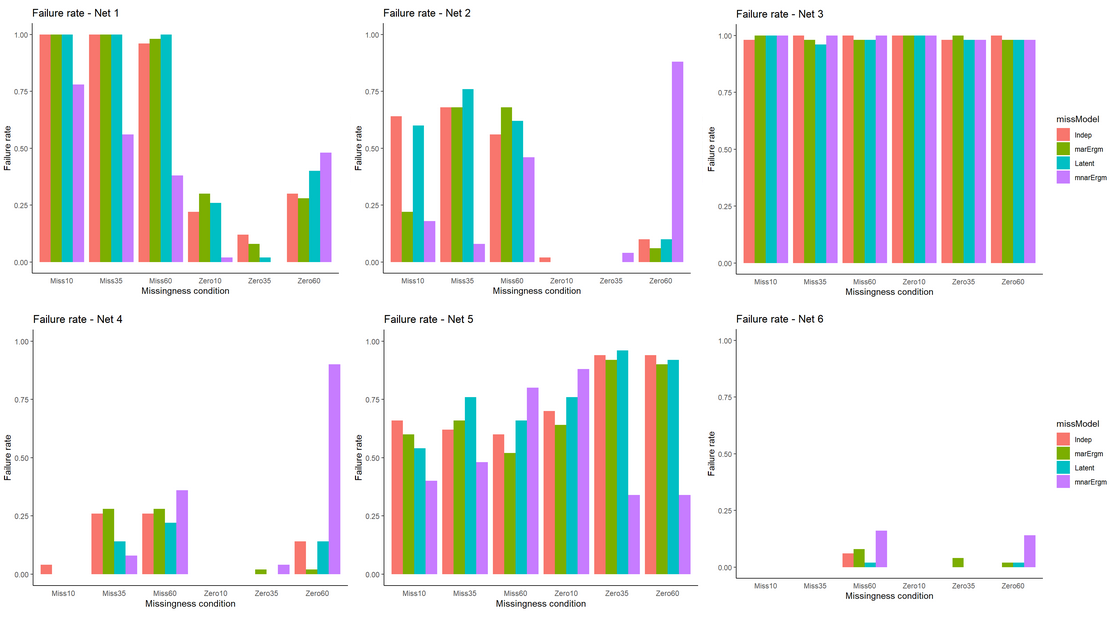}
    \caption{Six bar plots describing the failure rate for each empirical network. The x-axis represents different missingness conditions depending on how missingness was represented and the proportion of missing tie variables. The y-axis represents a failure rate as specified above. Differently coloured bars represented different missingness models. }
    \label{fig:failPlots}
\end{figure}

As seen in Figure \ref{fig:failPlots} the bar plots for Network 1 are notably higher when missingness was represented as NA values. In other words, the models are noticeably easier to estimate if the missingness were represented as 0 as the use of the face-value likelihood is circumvented. A similar tendency can be seen in the bar plots for Network 2, suggesting that the way missingness is represented can have practical consequences on model estimation. The bar plots for Network 3 suggest estimation difficulties for any missingness condition. The bar plots for Networks 5 also suggest estimation difficulties for many missingness condition. 

Lastly, the bar plots for Network 6 show that the estimations for Network 6 had the least difficulties despite similarities in its density to Network 5. We tentatively attribute the lack of difficulty to be the specification of a more elaborate estimation model as the estimation model for Network 6 contained attribute covariates in the model. Lastly, we can see a noticeable difference between the MNAR missingness model compared to the other three missingness models, possibly due to the algorithmic routine assuming MAR.

We find that the true network, above and beyond any kind of missingness, can affect model estimation severely. As seen with Network 3 and its corresponding failure rates, dense networks where some nodes have very high degree lead to severe estimation difficulties. This is a documented issue when dealing with ERGMs as max degree nodes in a degree distribution can lead to near degenerate distributions \parencite{handcock_assessing_2003}. In the face of many different conditions of missingness, any estimation model involving Network 3 had major degeneracy problems. It should also be noted that even without any missingness, Network 3 had notable estimation difficulties. Nonetheless, a model could still be estimated for Network 3. Any missingness that was introduced further exacerbated the estimation difficulties.

We interpret the lack of difficulty in Network 6 as the effect of the model specification. The choice of the model specification is well known to be a particularly important decision for network models \parencite{lusher_exponential_2013, snijders_markov_2002, handcock_assessing_2003}. Upon visual inspection of the network, Network 6 was not distinguishably different to the other networks. However, as seen from the low failure rates, the re-estimations involving Network 6 was least affected by missingness. 

\subsection{Relative bias}

A relative bias and relative standard error was used to capture the difference between the 'true' parameter estimate and the parameter estimated under a particular missingness condition. The difference of the estimate was captured with a relative bias parameter (rBias). This was computed by taking the difference of a chosen parameter for re-estimated model under a particular missingness condition  and the baseline model without any missingness. We captured spread by taking the ratio of the standard errors of the chosen parameter for the re-estimated model and the baseline model, which we defined as the relative standard error (rSe). Specifically, the relative metrics were defined as 

$$rBias = \frac{\tilde{\eta} - {\eta}}{\eta}, \ rSE = \frac{SE(\tilde{\eta})}{SE(\eta)}.$$

\begin{figure}[H]
    \centering

\includegraphics[width=10cm]{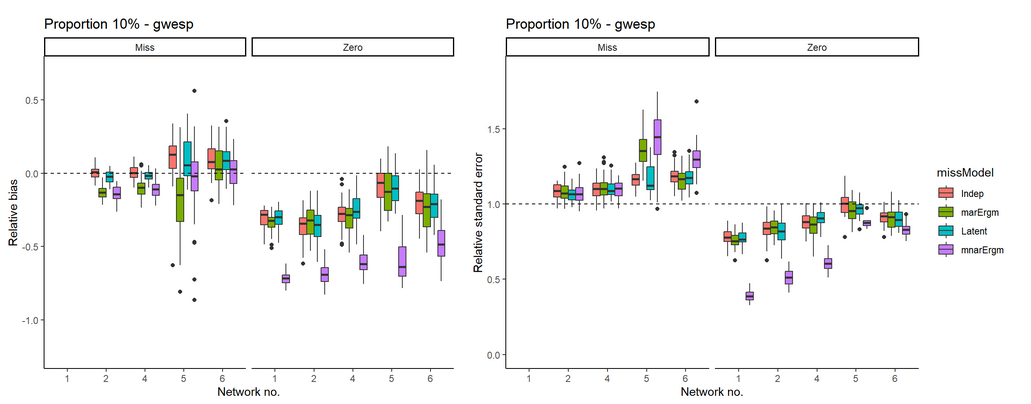}

\includegraphics[width=10cm]{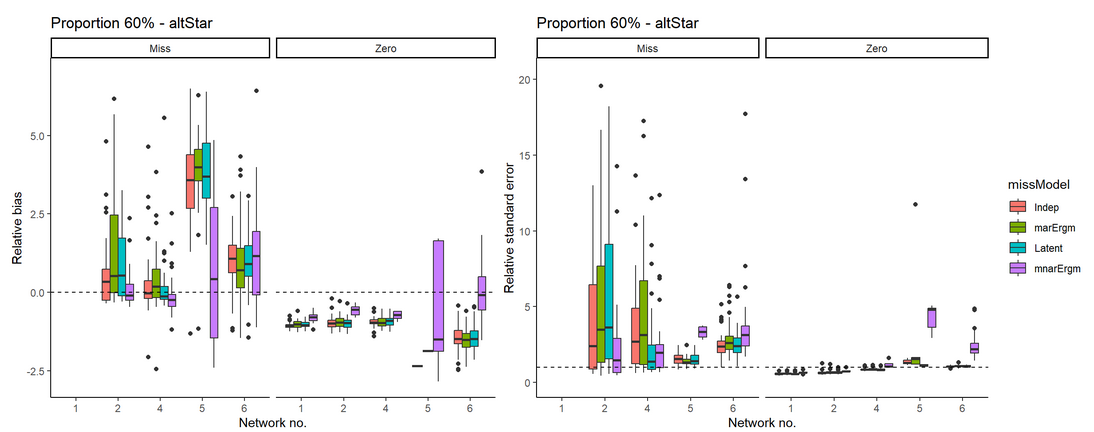}

    \caption{Each row refers to a specific estimation parameter, $\eta$, and proportion of missingness. The left column refers to the relative bias parameter and the right column refers to relative standard error parameter. The horizontal axis represents different networks while the vertical axis represents the parameter under evaluation. The left side of each plot refers to missingness when represented as missing while the right side refers to missingness represented as zeroes. Differently coloured boxplots refer to different missingness models. Note that Network 3 is absent in these plots due to model estimation problems.}
    \label{fig:relativeBoxplots}
\end{figure}

In the subset of results seen in Figure \ref{fig:relativeBoxplots}, we have the GWESP parameter under a small amount of missingness (10\%) for the top row and the alternating star parameter under a large amount of missingness (60\%) in the bottom row. This subset was chosen as they illustrated most of the consistent trends across the plots.  
Generally for both the M(C)AR and MNAR ERGMs, the GWESP parameter is underestimated for most of the networks. We also see the MNAR ERGMs lead to noticeably inflated standard errors in the Miss condition.

In contrast to the failure rate plots distinguishing Network 6 to the other networks, the relative bias plots involving Network 6 does not appear to be particularly robust to biased estimates. We also note that when the missingness proportion is large, the variance of the estimate can still be notably affected by the missingness. Therefore, we conclude that specification of the estimation model is tremendously helpful for the convergence of the estimation model, but is not robust to the biased estimates from the missingness mechanisms. We posit the true structures of the network and the extent to which they were depleted for each network to explain the heterogeneities in the relative bias plots.

On the topic of how missingness is represented (Zero vs Miss), we note some findings consistent with the expectations above. By removing the possibility of missing data points to be ties, the estimated coefficients intuitively tend towards 0 particularly when the proportion of missingness is high. The variance of the Miss representation can also be seen to be larger than the Zero representation. In other words, narrowing down the possible networks given the missing data artificially decreased uncertainty in the model. We note that the large variance in the Miss condition is to be expected from the face-likelihood framework \textcite{handcock_modeling_2010} and the computation of the Hessian as performed in statnet \parencite{krivitsky_likelihood-based_2023} and discussed above in Section 5.3.

Despite the MCAR and MNAR ERGMs implying different missingness assumptions, the relative bias of the GWESP parameter in the small missingness condition under the Miss condition were comparable. This was not expected as the heterogeneous ERGMs are nonetheless analytically MCAR and thus should be handled using the 'ergm' package routines. We note this as an illustration of the impact of having heterogeneity between missing tie variables on model estimation.

The examples above provide an interesting viewpoint for evaluation of MNAR mechanisms. We generally assume that the missingness is MNAR if the missingness were conditional in any way on the network. We used and varied the entrainment parameter above to evaluate the MNAR assumption in a relatively simple case. We generally noted MNAR missingness to have distinct effects compared to the other missingness models in terms of bias and variance. Our further simulations also noted that different MNAR specifications affected the estimation model differently. We did also note that simulation of the mean value parameters was a convenient and reasonably accurate method to assess the effects of different MNAR mechanisms. This method of assessment could potentially be useful in future research for evaluating different MNAR mechanisms. We demonstrate a brief example of assessing different MNAR mechanisms below.

\subsection{MNAR inspection}

In order to elaborate the flexibility afforded by different missingness model specifications provided by Equation \ref{equation:missGeneral}, we explore a few specific missingness mechanisms to clarity the extent of their effects. Specifically, we explore the \emph{entrainment} ($\theta_1$) and \emph{degree covariate} ($\theta_2$) parameters. 

As entrainment captures the relationship between the tie variable and its corresponding missingness indicator, different levels of entrainment changes whether ties or null ties are more likely to be missing. Specifically, a positive entrainment parameter assumes that ties are more likely to be missing while a negative entrainment parameter assumes that null ties are more likely to be missing. As the entrainment parameter weighs the extent to which the missingness indicator depends on the true data, it can be considered to weigh the MNAR assumption with a 0 entrainment parameter reflecting an MCAR assumption as the missingness model is no longer conditional on the data. 

In these sets of simulations, we chose the Jemaah Islamiyah 2005 Bali bombing network, Network 4, and degraded the network with 50 different realisations of five different levels of entrainment and evaluated some metrics. The five different levels of entrainment were -1, -0.5, 0, 0.5, and 1 to reflect higher and moderate levels of missingnss for both ties and null ties, and 0 to reflect an MCAR assumption primarily as a reference. All other parameters in the missingness model was set to 0 and the proportion of missingness was fixed to 35\% to narrow down the effect from different entrainment levels as seen in Table \ref{table:mnarInspection}. 

\begin{table}[H]
\begin{tabular}{llll}
\hline
Parameters  & Statistics      & \multicolumn{2}{c}{Missingness model}          \\ \hline
              &    & Entrainment & Degree covariate \\
& Edges   $\sum_{i < j} d_{ij}$         & Fixed         & Fixed                            \\
$\psi_2$ & GWDegree   $e^{\alpha}\sum_{k = 1}^{n-1} \{1 - (1 - e^{-\alpha})^k\}D_k(d)$      & 0         & 0        \\
$\psi_3$ & GWESP    $e^{\alpha} \sum_{i = 1}^{n-2} \{1 - (1 - e^{-\alpha})^i \} sp_i$        & 0       & 0                    \\
$\theta_1$ & Entrainment   $\sum_{i < j} d_{ij} x_{ij}$              & Varying       & 0                \\
$\theta_2$ & Degree covariate $\sum_{i < j} d_{ij} \sum_{i} x_{ij}$ &         0    & Varying                      \\ \hline
\end{tabular}

\caption{Table describing the simulated missingness models for the follow-up MNAR inspection}
\label{table:mnarInspection}

\end{table}

We then degraded the network using the 50 realisations of each of the 5 levels and re-estimated the degraded networks. For reference purposes, mean value parameters are reported for zero imputation (i.e., $\hat{\mu}_{Zero}(s(\mathbf{Y}) ; \mathbf{D})=s( \mathbf{X}_{obs},\mathbf{X}_{mis}=\mathbf{0})$). The mean value estimates under the MAR assumption could have been obtained as $\mu_{\hat{\eta}}(s(\mathbf{Y}) \mid \mathbf{X}_{obs})$ but are more computationally demanding.
Similarly,  other metrics were likewise simulated by assuming the missing values were 0.

\begin{figure}[H]
    \begin{subfigure}{0.42\textwidth}
        \centering
        \includegraphics[height = 5cm, width=\linewidth]{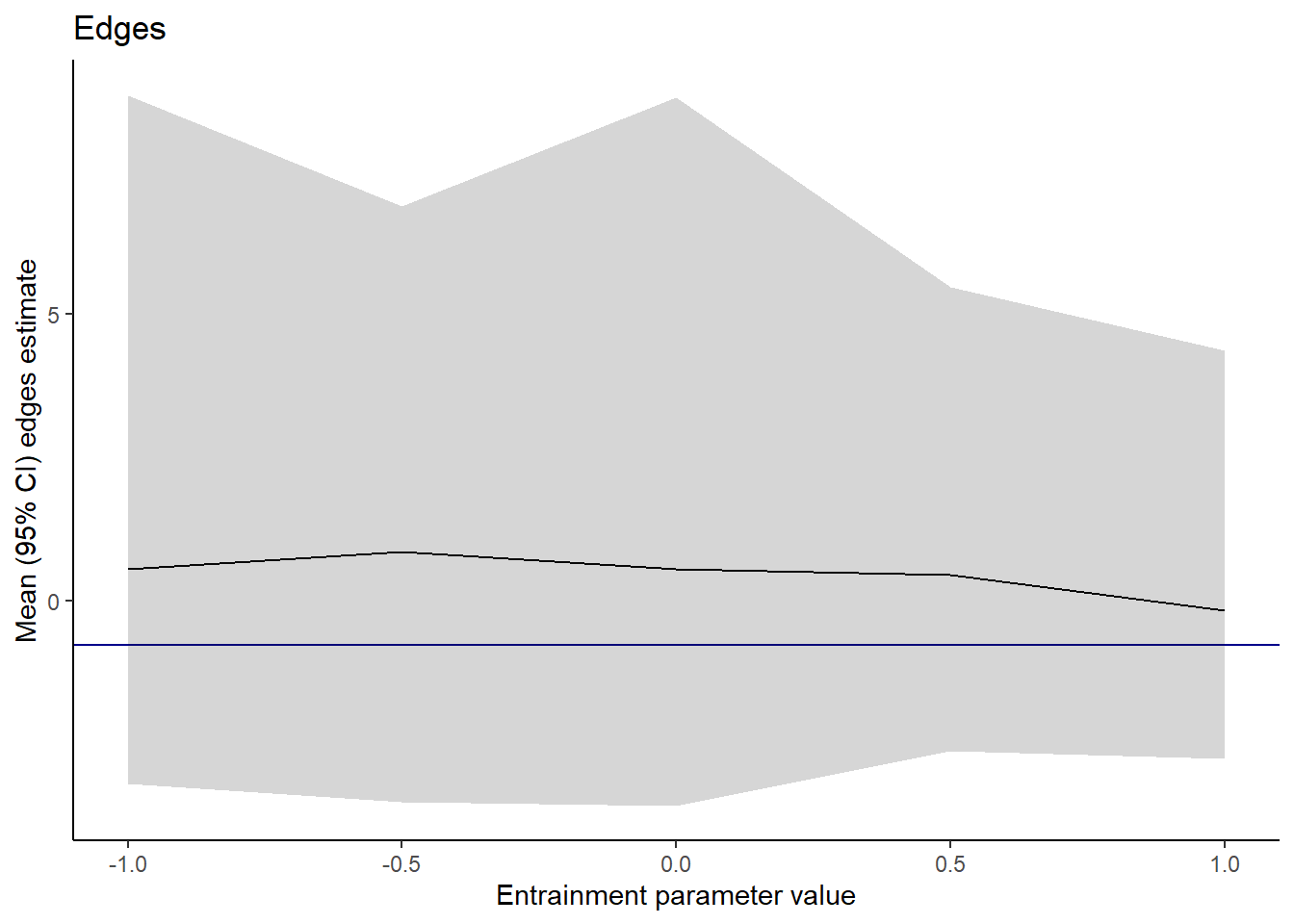}
        \caption{Density estimate $\hat{\eta_1}$}
        \label{fig:entrSimEdgeEst}
    \end{subfigure}
    ~
    \begin{subfigure}{0.42\textwidth}
        \centering
        \includegraphics[height=5cm, width=\linewidth]{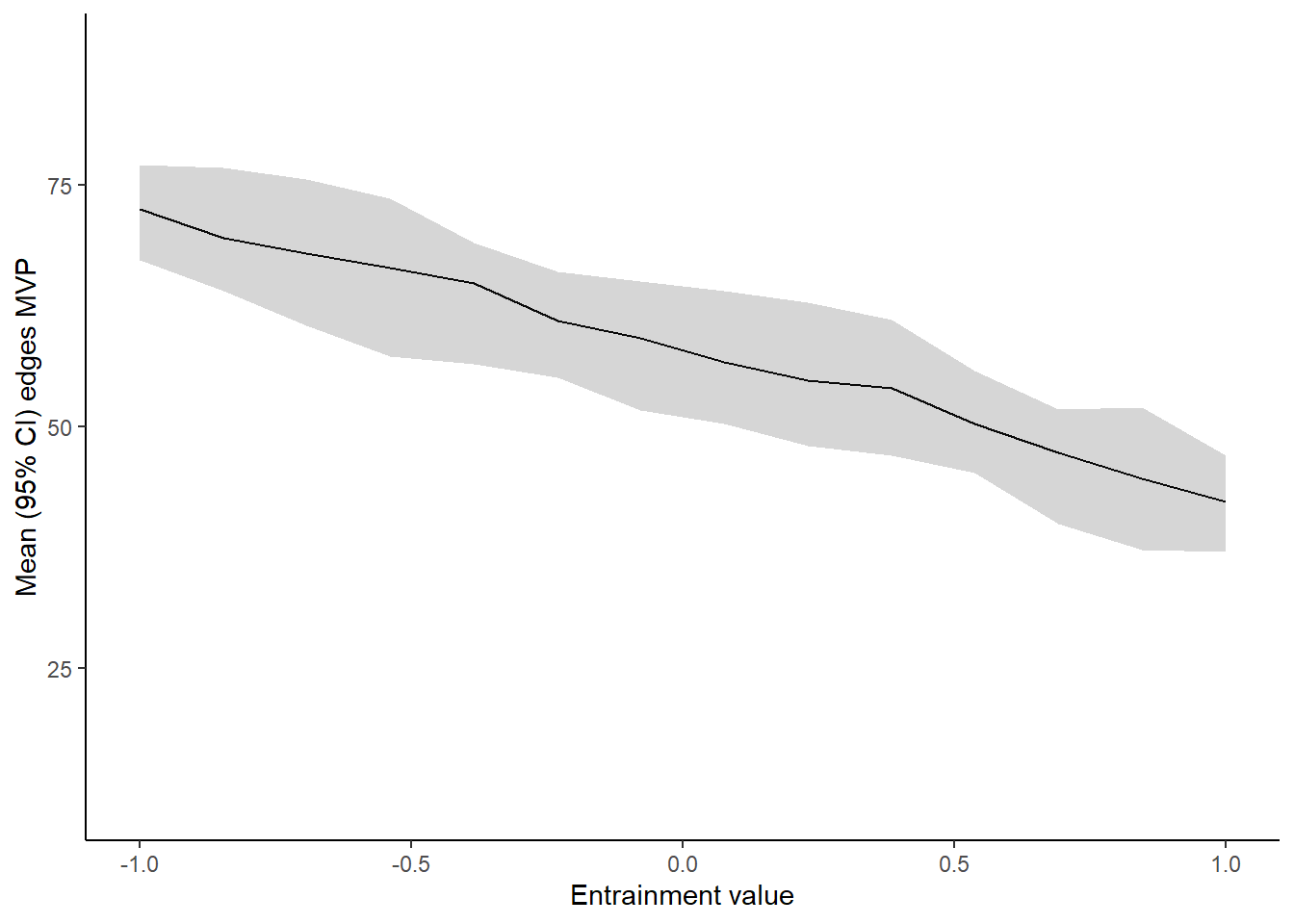}
        \caption{Observed edge count}
        \label{fig:entrSimEdgeMvp}
    \end{subfigure}
    \caption{The lines refer to the mean of the re-estimated parameters, $\hat{\eta}$ (left-hand panels), and simulated parameters $\hat{\mu}_{Zero}(s(\mathbf{Y}) ; \mathbf{D})$ (right-hand panels). The shaded area refer to the 95\% probability intervals from the parameters of the simulated sampling distributions. The dark blue line refers to the true estimate for the completely observed data for reference purposes.}
    \label{fig:entrPlotEdge}
\end{figure}

The exploration of the effects of the entrainment parameter $\theta_1$ clearly demonstrate the systematic effects of MNAR mechanisms. From Figure \ref{fig:entrPlotEdge}, we see that the density parameter $\hat{\eta_1}$ (left-hand panel) is unexpectedly stable despite differences in $\theta_1$ values. It is especially remarkable because the observed count of edges ($\hat{\mu}_{Zero}(s(\mathbf{Y}) ; \mathbf{D})$) are clearly decreasing (right-hand panel). 

\begin{figure}[H]
    \begin{subfigure}{0.42\textwidth}
        \centering
        \includegraphics[height = 5cm, width=\linewidth]{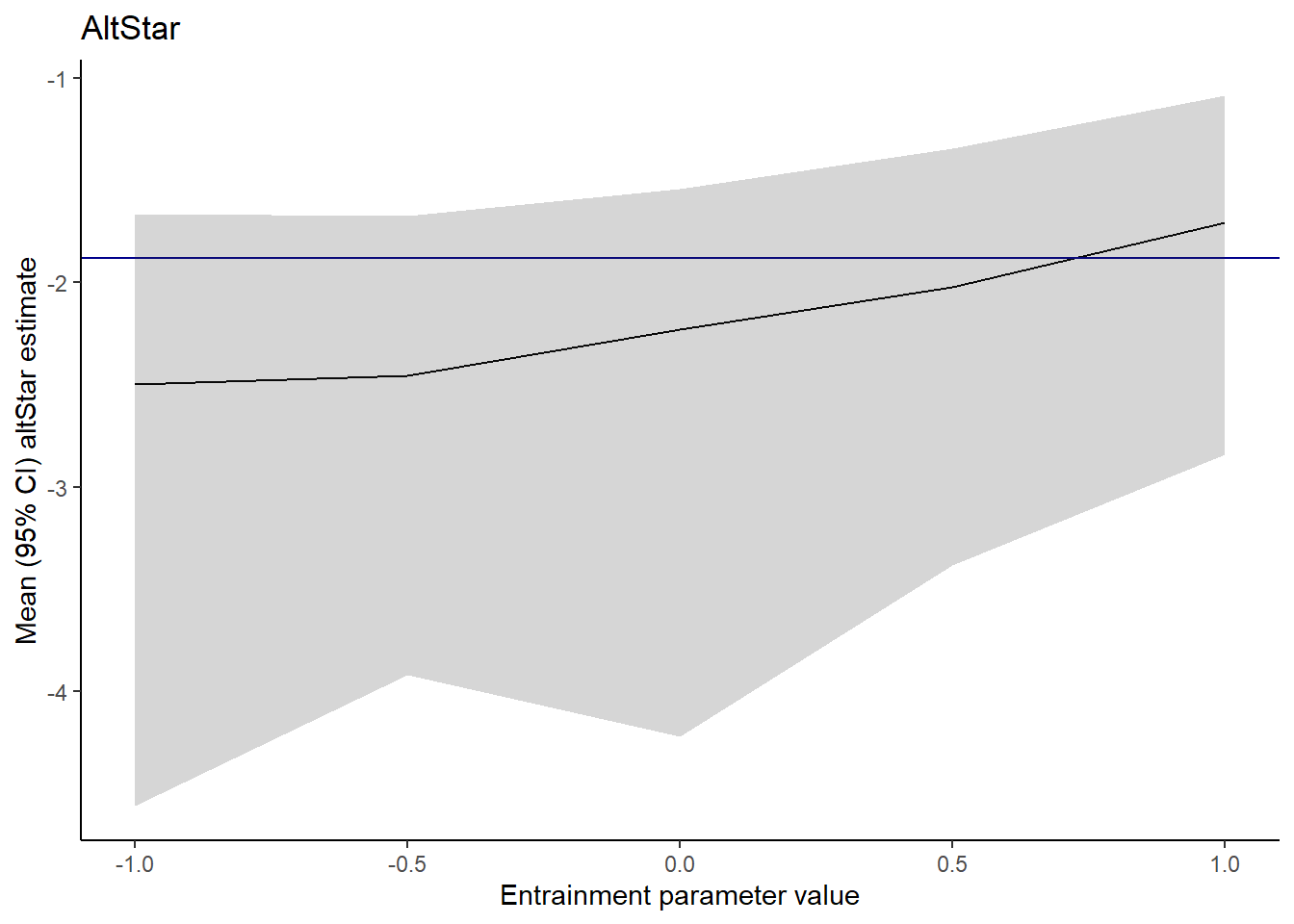}
        \caption{Alternating star estimate $\hat{\eta_2}$}
        \label{fig:entrSimAltStarEst}
    \end{subfigure}
    ~
    \begin{subfigure}{0.42\textwidth}
        \centering
        \includegraphics[height=5cm, width=\linewidth]{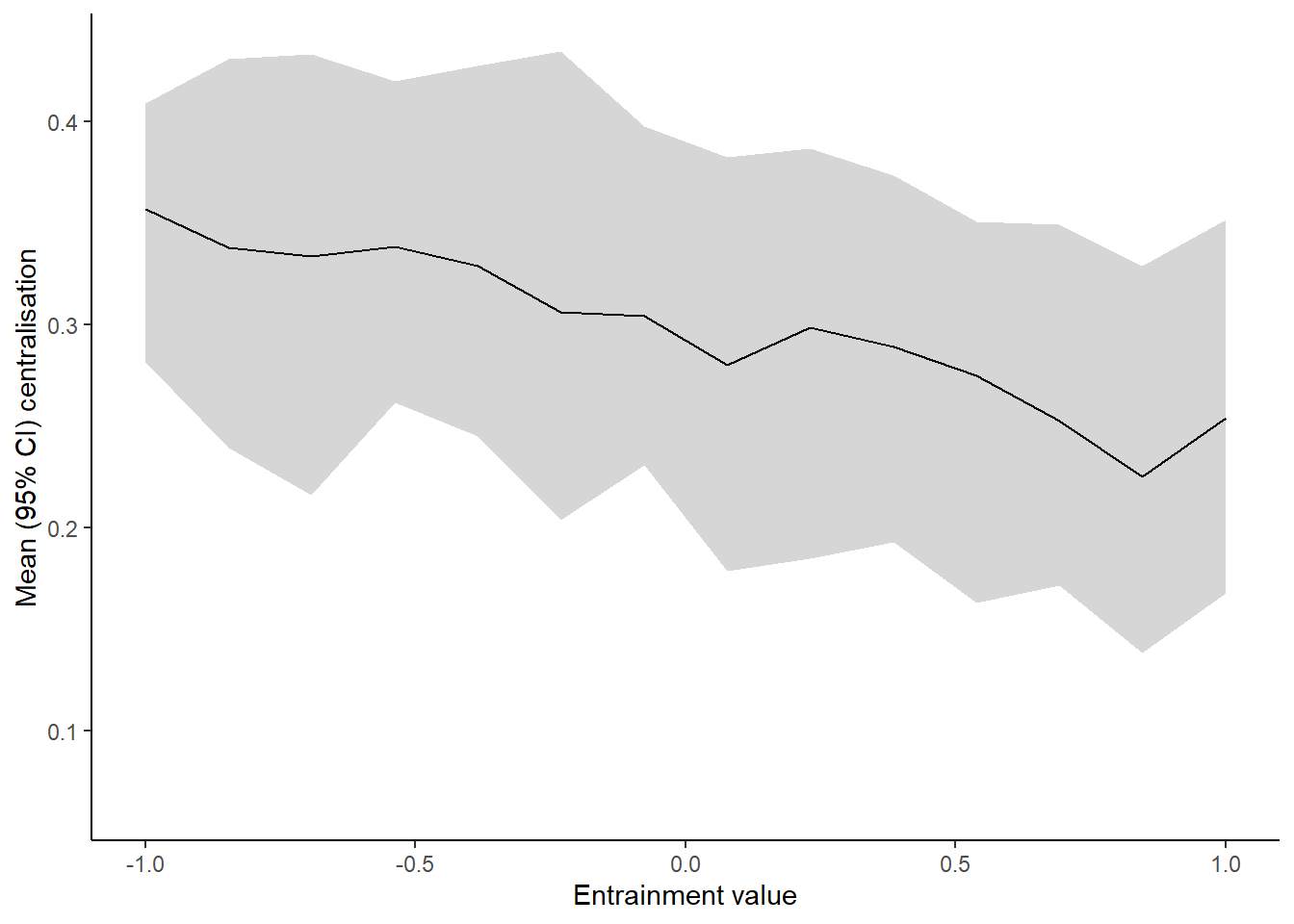}
        \caption{Centralisation metric}
        \label{fig:entrSimCentr}
    \end{subfigure}
    \caption{Identical plotting specifications to Figure \ref{fig:entrPlotEdge}}
    \label{fig:entrPlotCentr}
\end{figure}

The unexpected stability of the density parameter $\hat{\eta_1}$ can be explained by Figure \ref{fig:entrPlotCentr}. Here, we see that the centralisation metric (right-hand panel) decreases as the number of edges decreases. However, the alternating star parameter, $\hat{\eta_2}$, which captures the centralisation metric, increases. We can understand the stability of the density parameter $\hat{\eta_1}$ to be a consequence of the inclusion of $\hat{\eta_2}$ as the alternating star parameter inherently accounts for the network's degree distribution. 

\begin{figure}[H]
    \begin{subfigure}{0.42\textwidth}
        \centering
        \includegraphics[height = 5cm, width=\linewidth]{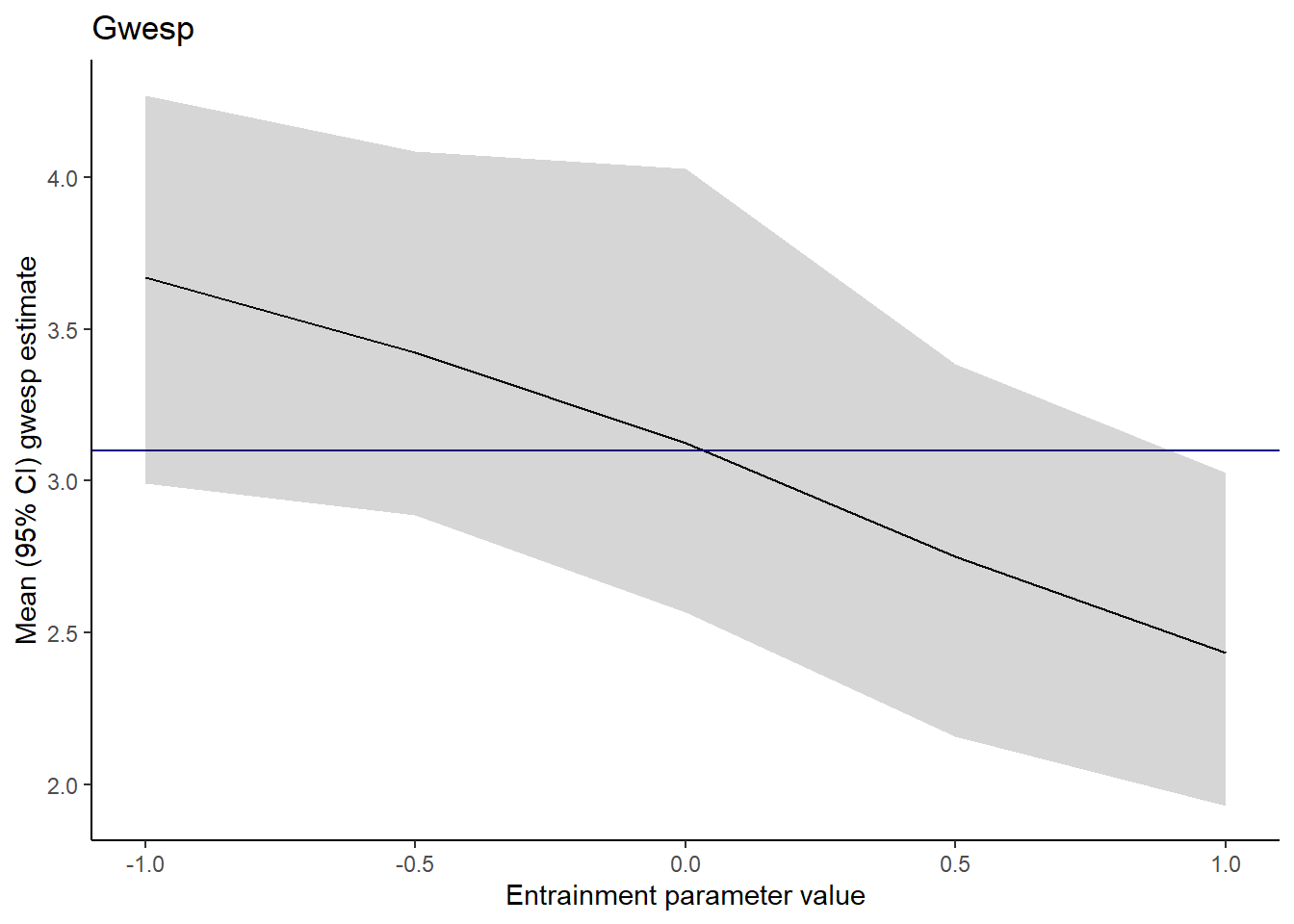}
        \caption{GWESP estimate $\hat{\eta_3}$}
        \label{fig:entrSimGwespEst}
    \end{subfigure}
    ~
    \begin{subfigure}{0.42\textwidth}
        \centering
        \includegraphics[height=5cm, width=\linewidth]{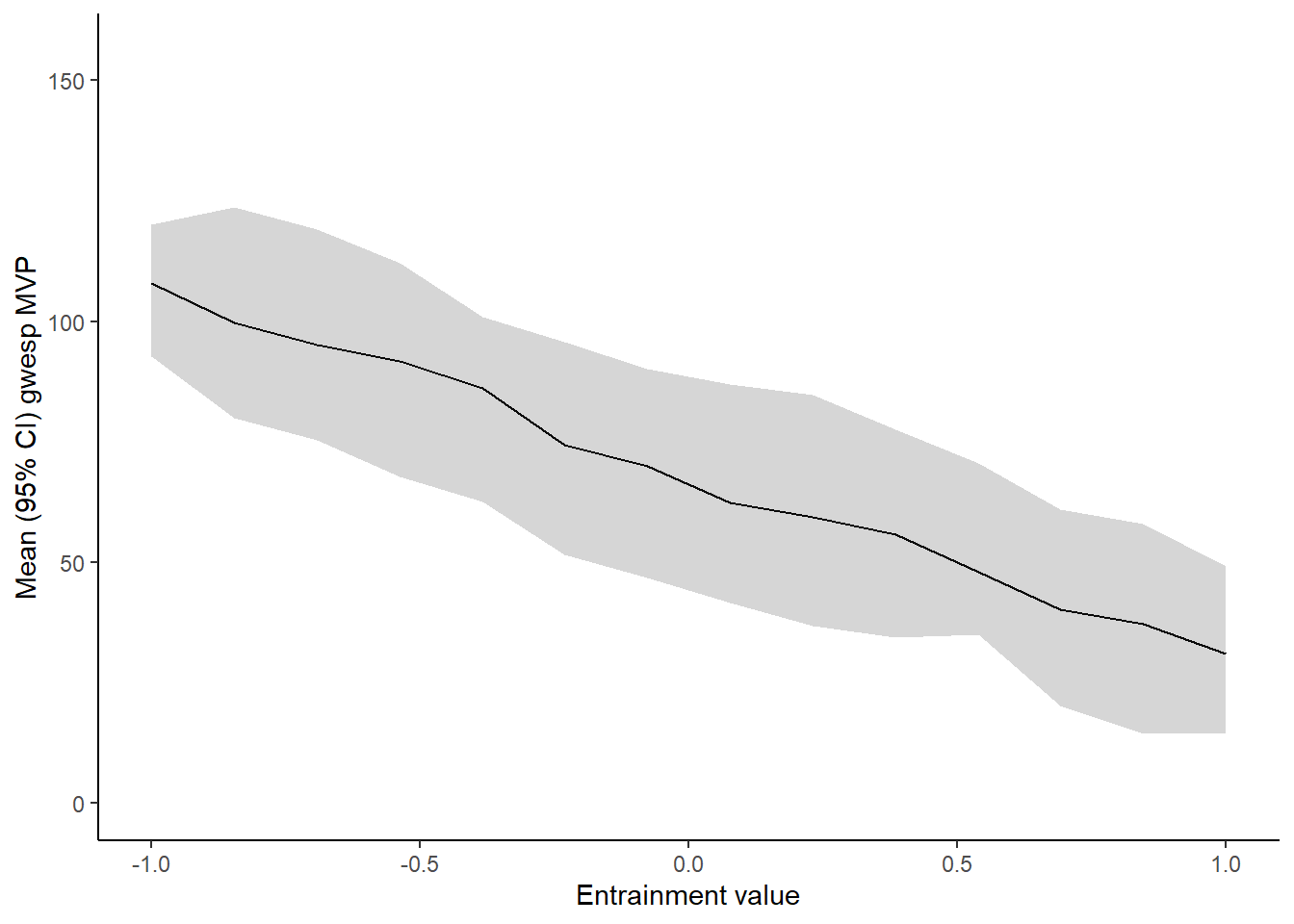}
        \caption{GWESP parameter count}
        \label{fig:entrSimGwespMvp}
    \end{subfigure}
    \caption{Identical plotting specifications to Figure \ref{fig:entrPlotEdge}}
    \label{fig:entrPlotGwesp}
\end{figure}

We see the strongest effects of the entrainment parameter in Figure \ref{fig:entrPlotGwesp}. We clearly see that the GWESP parameter $\hat{\eta_3}$ (left-hand panel), is clearly very sensitive to the strength and sign of $\theta_1$. At $\theta_1 = 0$, which implies MAR, we see that the sampling distributions are centered on the 'true' target value. However, through our simulations we can see that the triadic dependence can be severely under or over-estimated depending on the value of the entrainment. This may seem trivially a consequence of the decreasing GWESP statistic (right-hand panel). However, as the sole missingness mechanism manipulated was the entrainment parameter $\theta_1$, we can understand the decrease in the clustered triadic dependence \emph{as a consequence of a missingness mechanism targeting edges}.

In conclusion, these set of simulations demonstrate the impact of specific missingness mechanisms, and consequently the utility of evaluating them through the ERGM. We find that the centralisation metric is fairly robust to missing edges. We also find that inference involving clustering-based metrics are not robust to missing edges as demonstrated by the entrainment parameter $\theta_1$.

\begin{figure}[H]
    \begin{subfigure}{0.42\textwidth}
        \centering
        \includegraphics[height = 5cm, width=\linewidth]{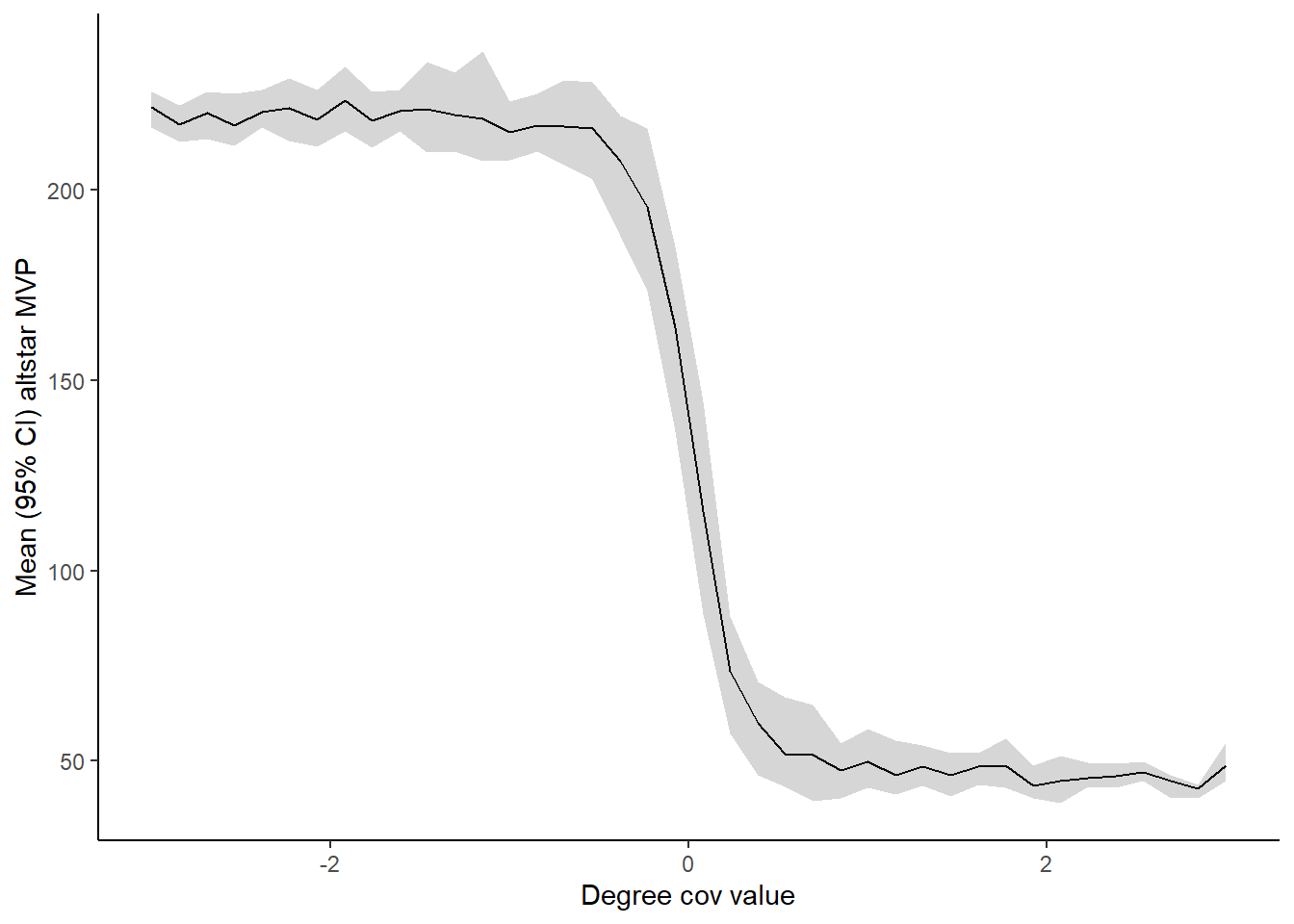}
        \caption{Alternating star counts}
        \label{fig:degCovSimAltStarMvp}
    \end{subfigure}
    ~
    \begin{subfigure}{0.42\textwidth}
        \centering
        \includegraphics[height=5cm, width=\linewidth]{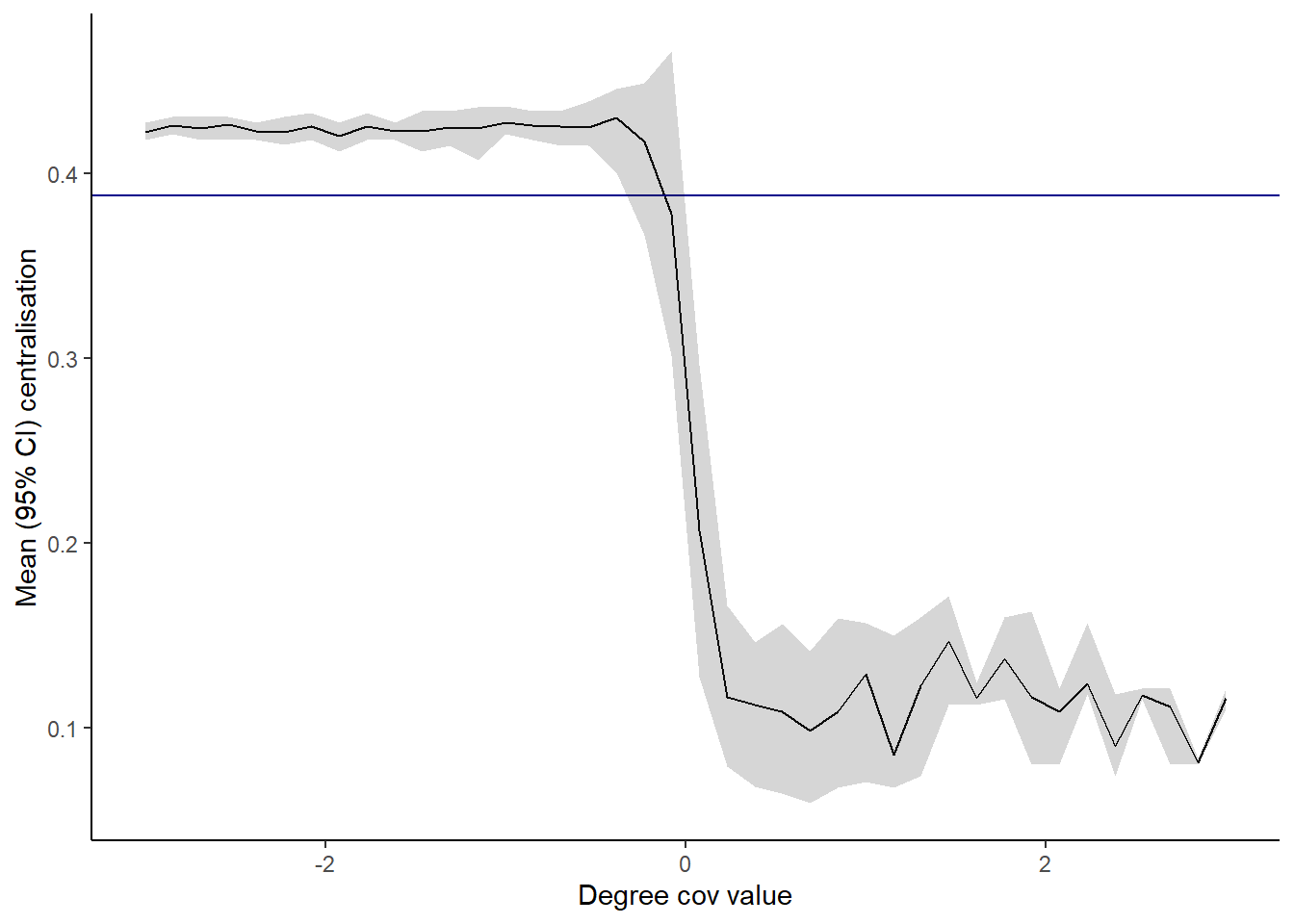}
        \caption{Centralisation metric}
        \label{fig:degCovSimCentr}
    \end{subfigure}
    \caption{The effect on centralisation by $\theta_2$. The left plot refers to the simulated alternating star mean value parameter estimates and the right plot refers to the global centralisation metric. The shaded area refers to the 95\% confidence interval. A dark blue reference line was included to refer to the true network's value.}
    \label{fig:degCovPlots}
\end{figure}

As seen in Figure \ref{fig:degCovPlots}, varying $\theta_2$ has a much more marked effect on the network inference than $\theta_1$. A phase-transition in \emph{observed} network centralisation can be seen, with a sharp drop from high centralisation to low centralisation around $\theta_2= 0 $. The regimes either side of the MCAR condition $\theta_2 = 0$, are remarkably stable.

The further simulations examining different specifications of the MNAR model also show that there are differences in the consequences of MNAR missingness models depending on how the MNAR missingness is specified. By comparing the entrainment and degree covariate plots, we note how different values of both parameters can have radically different effects on the observed network, over and above changing the missingness assumption (Table \ref{table:missAssumps}). While entrainment $\theta_1$, seemed to have more of a linear effect on the observed metrics, the degree covariate parameter $\theta_2$ had a strongly non-linear effect with phase transitions.

The similarities between the changes in the re-estimated parameters $\hat{\eta}$ (e.g. the left-hand panels of Figures \ref{fig:entrPlotEdge}-\ref{fig:entrPlotGwesp}) for different entrainment values $\theta_1$, and parameter values $\hat{\mu}_{Zero}(s(\mathbf{Y}) ; \mathbf{D})$ under the Zero condition (e.g. the right-hand panels of Figures \ref{fig:entrPlotEdge}-\ref{fig:entrPlotGwesp}) suggest that simulations under the Zero condition can be a useful tool to quickly evaluate the consequences of missingness model specifications. There are, however, some differences as seen in the simulated edges parameter compared to the re-estimated edges coefficient. This can be explained through the different algorithms in place for handling missing observations. Differences in the simulation and re-estimation can arise due to preexisting software routines to handle missing data. In summary, representing missingness through a specified missingness model can assess a range of assumptions about the missingness process and their consequences can be reflected in multiple ways.

\section{Discussion}

In this paper we have presented a framework to model missingness through various statistical network models and investigated their effects on empirical covert network data. In particular, we suggest that exponential random graph models (ERGMs) parameterise different, distinct, data collection biases.
Our illustrative examples demonstrate different empirical biases in the ways missing tie variables can occur and their consequences for analysis. We emulate and evaluate the scenario of a network analyst using current gold standard methods and software to analyse partially observed networks. In our examples, we demonstrate how model degeneracy problems can be exacerbated by the missingness. We also illustrate the effects of implicit algorithmic routines on the estimation of partially observed networks.

In lieu of the examples from the failure rates and relative bias plots, we highlight two important concepts. The first is of the different true network structures. As depicted by the failure rate plots, the structure of the true underlying network can affect the rate of successful model estimations. The failure rates can however be improved by the second important concept, model specification. Model specification of the ERGM has historically been important for inferential and computational reasons \parencite{handcock_assessing_2003, snijders_markov_2002, snijders_new_2006} and we echo the sentiment in our examples. As the model specified for Network 6 was more elaborate and contained more covariates, the failure rates were remarkably small. However, the estimates from a well specified model is not robust to the biases introduced by various missingness mechanism. It is the true network structures and the extent to which they are depleted that can explain the biases. Therefore, while model specification is important in the convergence of the estimation model, it is the true network structures that have the greatest effects on the relative biases induced by various missingness mechanisms.

In our comparison of missingness and dependence assumptions, we note how the standard MAR assumption cannot be motivated through endogenous dependence between the missing tie variables. Outside of the case where an independent covariate, or a set of covariates, can explain all aspects of the missingness mechanism, we note how dependence assumptions within the missingness mechanism is likely to imply either MCAR or MNAR. Previously it has been shown how most statistical network models admit the seemingly general MAR assumption, but, outside of snowball sampling, implied missingness will more likely be MCAR - which is implausible - or MNAR - something that current estimation techniques cannot deal with. 
In our formulation of a general case for the missingness model (Equation \ref{equation:missGeneral}), we clarified how different model parameters can affect the missingness assumption.

When we narrow down the scope of our analyses and narrow down specific missingness mechanisms, we find MNAR mechanisms ($\theta_1$ and $\theta_2$) to severely attenuate inference for clustering. MNAR mechanisms also inflate the uncertainty in the estimation and we demonstrate the utility of sensitivity analyses. From a closer look at the models where ties are more ($\theta_1>0$) or less ($\theta_1<0$) likely to be missing, we see that inference for clustering (GWESP) is strongly affected by the dependence on the true network. Therefore, if ties are more (or less) visible than null ties, but you analyse the network assuming MAR, you will severely underestimate (or overestimate) triadic closure in covert networks. When we allow the true (but unknown) centrality of an actor to affect how likely it is that their ties are observed, we find a dramatic phase-transition as a function of the strength of this dependency ($\theta_2$).

Beyond the specifics of model estimation and mathematical implications, we note a few pointers for the network analysts and the broad covert network community. The examples have demonstrated the utility of covariate data (e.g., age, group affiliation) and we would like to emphasise their collection. We also argue that covariate data would be easier to triangulate across different sources of information when compared to different tie definitions. In the effort to reduce ambiguities arising from composite tie definitions, we also point towards strongly considering the edge definition when defining ties in the network.

As seen in the examples, a denser network can lead to difficulties in model estimation. It is understandable that the investigator and/or analyst would want to collect as much data as they can, however we propose a more guided approach. Establishing a framework to clarify which relationships are being examined (e.g., familial, transactional, business, etc) and how they overlap (e.g., family businesses,etc.) will be beneficial when making substantive claims about the network (e.g., 'family members are more likely to be a part of the covert network'). It is admittedly counterintuitive to restrict the already-scarce sources of relational information. However evidently seen in Network 3 (Jemaah Islamiyah embassy bombing), having more relational information is not always beneficial in a modelling perspective. On the substantive side, generally knowing two individuals are somehow related adds further ambiguity and obscures membership of the covert network.

An exhaustive list of plausible missingness mechanisms is beyond the scope of this paper, much less a description of which ones should be considered to be in operation for any particular research scenario. Instead, we have explored the concept of explicit missingness models and demonstrated a few examples of their use in the representation of complicated missingness mechanisms. The ERGM for missing data, provides the researcher with a convenient way of quantifying how missingness can beget missingness, of quantifying active cloaking of illicit interaction, of studying the influence of the spotlight effect, etc. 

A natural next step, having developed an explicit missingness model, is to use it in sensitivity analysis and imputation. As the missingness is described as a statistical model, we could incorporate the ERGM missingness model with the ERGM for data with Bayesian data augmentation in \textcite{koskinen_bayesian_2013} to extend their algorithm to more complicated missingness assumptions. We would also like to restate that the method of using missingness models is more general than covert networks and may be explored in the context of survey-collected data where missingness induced by non-respondents can be addressed.

\section{Data availability}
The study uses publicly available data as indicated and referenced in the text. All figures and code are available at request from the corresponding author (J. Januar).

\section{CRediT authorship contribution statement}
\textbf{Jonathan Januar}: Conceptualisation, Formal analysis, Methodology, Software, Visualisation, Writing - original draft, Writing - review \& editing. \textbf{H Colin Gallagher}: Conceptualisation, Methodology, Resources, Writing - review \& editing, Supervision. \textbf{Johan Koskinen}: Conceptualisation, Methodology, Software, Writing - review \& editing, Supervision, Funding acquisition.

\section{Acknowledgements}
This project was financially supported by the project `Covert Networks: How to learn as much as possible about the structure of a network from sampled subnetworks' funded by Department of Defence, through the US Army Research Office (Grant Number  W911NF-21-1-0335 Proposal 79034-NS). Januar gratefully acknowledges input and comments on early drafts from Garry Robins. The authors declare no conflicts of interest.

\emergencystretch=1em

\printbibliography

\appendix

\section{More MAR examples}

If we were to instead assume that the missingness of a tie variable is evaluated independently with respect to its corresponding tie variable with MAR we get

$$g_{11}(x_{ij}, x_{ik}, \psi) = g_{1+}(x_{ij}, \psi) g_{+1}( x_{ik}, \psi).$$

\noindent Where $g_{1+}(x_{ij}, \psi)$ is defined as

$$g_{1+}(x_{ij}, \psi) = g_{11}(x_{ij}, \psi) + g_{10}(x_{ij}, \psi),$$

\noindent and $g_{+1}(x_{ik}, \psi)$ is defined as

$$g_{+1}(x_{ik}, \psi) = g_{11}(x_{ik}, \psi) + g_{01}(x_{ik}, \psi).$$

\noindent In both expressions, we can see that the missingness of the tie variable is dependent only on itself, which would be MNAR with the exception of MCAR if $g_{1+}(x_{ij}, \psi) = g_{1+}(\psi).$
This suggests that we cannot realistically assume MAR for individual tie variables. If we extend the independence of missingness in a single variable to independence in a bivariate case, this yields

$$g_{11}(x_{ij}, x_{ik}, \psi) = g_{1+}(x_{ij}, \psi) g_{+1}(x_{ik}, \psi),$$
$$g_{10}(x_{ij}, x_{ik}, \psi) = g_{1+}(x_{ij}, \psi) (1 -  g_{+1}(x_{ik}, \psi)),$$
$$g_{01}(x_{ij}, x_{ik}, \psi) = (1 - g_{1+}(x_{ij}, \psi)) g_{+1}(x_{ik}, \psi),$$
$$g_{00}(x_{ij}, x_{ik}, \psi) = (1 - g_{1+}(x_{ij}, \psi))(1 -  g_{+1}(x_{ik}, \psi)).$$

\noindent We can see that MAR is not satisfied as the probability of one tie variable being observed requires knowledge about the probability of the other tie variable being missing. As the missingness of $x_{ij}$ depends on the missing tie variable $x_{ik}$, this example is MNAR. In summary, when thinking about missingness and dependence assumptions, we can assume MCAR if we assumed that the true network does not affect missingness. Conversely, when we assume the true network to affect the missingness, MNAR would be the more likely case and MAR is comparatively difficult to substantively motivate. However, we have only addressed evaluating the missingness with information from the network and no external sources of information (e.g., covariates).

\end{document}